%% file: kn_arxiv.tex
\documentclass[twocolumn,twocolappendix,trackchanges]{aastex631}

\usepackage{amsfonts,amsmath,graphicx,natbib,nicefrac,url,hyperref,breakurl,
graphicx,lipsum,xcolor}
\usepackage{comment}

\input{defs}

\turnoffedit

\begin{document}

\title{Klein-Nishina Corrections to the Spectra and Light Curves of Gamma-ray Burst Afterglows}
\author[0009-0006-2557-6569]{George A. McCarthy}
\affil{Department of Physics, University of Bath, Claverton Down, Bath, BA2 7AY, United Kingdom}
\author[0000-0003-1792-2338]{Tanmoy Laskar}
\affiliation{Department of Physics \& Astronomy, University of Utah, Salt Lake City, UT 84112, USA}
\affiliation{Department of Astrophysics/IMAPP, Radboud University, P.O. Box 9010, 6500 GL, Nijmegen, The Netherlands}
\affil{Department of Physics, University of Bath, Claverton Down, Bath, BA2 7AY, United Kingdom}

\begin{abstract}
Multi-wavelength modeling of the synchrotron radiation from relativistic transients such as Gamma-ray Burst (GRB) afterglows is a powerful means of exploring the physics of relativistic shocks and of deriving properties of the explosion, such as the kinetic energy of the associated relativistic outflows. Capturing the location and evolution of the synchrotron cooling break is critical to break parameter degeneracies associated with such modeling. However, the shape of the spectrum above the cooling break, as well as the location and evolution of the break itself can be significantly altered by synchrotron self-Compton (SSC) cooling. We present an observer's guide to applying SSC cooling with and without Klein-Nishina (KN) corrections to GRB afterglow modeling. We provide a publicly available python code to calculate the Compton $Y$-parameter as a function of electron Lorentz factor, from which we compute changes to the electron distribution, along with KN-corrected afterglow spectra and light curves. In this framework, the canonical synchrotron spectral shapes split into multiple sub-regimes. We summarize each new spectral shape and describe its observational significance. We discuss how KN corrections can account for harder spectra and shallower decline rates observed in some GRB X-ray afterglows. Our overall aim is to provide an easy application of SSC+KN corrections into analytical multi-wavelength modeling frameworks for relativistic transients. 
\end{abstract}

\section{Introduction}
In the fireball model of $\gamma$-ray bursts (GRBs), their luminous afterglows are the result of synchrotron radiation produced in the relativistic forward shock (FS) formed in the interaction of their powerful jets with the ambient environment \citep{pr93,mr93,mr97,spn98,gps99,gs02}. Multi-wavelength modeling of GRB afterglows provides constraints on the physical properties of these extremely energetic explosions, including the isotropic-equivalent energy of the jet ($\EKiso$), the density of the ambient medium (parameterized as $\Astar$ for a wind-like medium with a density profile, $\rho\propto r^{-k}$ with $k=2$, or as the particle density, $\dens$, for a uniform-density (``ISM'') medium with $k=0$), the jet collimation (\thetajet), and the fraction of the FS energy given to magnetic fields ($\epsb$) and to relativistic electrons ($\epse$) that have been accelerated by the FS into a power-law distribution in Lorentz factor, $\frac{dn_0}{d\gammae}\propto\gammae^{-p}$ \citep[e.g.,][]{wg99,pk01,pk02,gs02, yhsf03,cfh+10,cfh+11,llw13,lbc+18,kf21,awvevdh22,slf+22}. Measurement of these properties, in turn, is essential for exploring particle acceleration in astrophysical shocks \citep[e.g.,][]{bo78,dru83,ew05,spi08,ss09,ss11,bgk+12,rl17,webn17}, in deriving clues to GRB progenitors \citep[e.g.,][]{cl99,cl00,lw00,pmn+04,bl06,wh06,knj08}, and for probing the central engine that is responsible for launching and collimating these powerful, transient, relativistic jets \citep[e.g.,][]{frw99,zkm03,tmn08,knj08a,kvkb09,lzl13,msw+13,bt16}.  

Multi-wavelength modeling of GRB afterglows usually relies on fitting the observed light curves and spectral energy distributions (SEDs) to the synchrotron model, wherein the received radiation is characterized by spectral power-law segments (PLSs) connected at break frequencies \citep{spn98}, which are, nominally, the synchrotron self-absorption (SSA) frequency ($\nusa$), the injection frequency (\numax), and the cooling frequency (\nuc). Locating each break provides novel physical constraints; e.g., $\nusa$ is strongly sensitive to the density, $\numax$ to $\epse$, and $\nuc$ to both $\epsb$ and the density. Capturing the cooling break is particularly valuable because the evolution of \nuc\ probes the density profile of the ambient medium ($\nuc\propto t^{-1/2}$ in an ISM environment and $\nuc\propto t^{1/2}$ in a wind medium; \citealt{spn98,cl00}). Furthermore, the flux density above the cooling break is independent of the density, a fact that has been exploited to devise a precision probe of \EKiso\ \citep[e.g.][]{fw01,wgmw16}. 

The evolution of $\nuc$ is affected by both synchrotron cooling and inverse-Compton (IC) losses. In particular, IC up-scattering of synchrotron photons by synchrotron-emitting electrons (synchrotron self-Compton, or SSC) can be an important cooling mechanism when the Compton $Y$-parameter is large (typically in regimes with $\epse/\epsb\gtrsim1$; \citealt{snp96,dcm00,pk00,se01,zm01}), and is expected to usually dominate over external inverse Compton \citep{zcp+20}. Incorporating SSC cooling when computing $\EKiso$ from afterglow X-ray light curves frequently indicates low post-shock magnetization ($\epsb\sim10^{-4}$), alleviating the problem of high prompt efficiencies \citep{kb09,bnp16}. Low values of $\epsb$ (implying strong SSC cooling) have also been found in moderately large samples of GRB X-ray and optical afterglows (e.g., \citealt{wll13, bdur14, sbdk14}), further highlighting the importance of properly accounting for SSC cooling in afterglow modeling. 
In addition to modulating $\nuc$, the inclusion of SSC effects in GRB afterglow modeling results in degeneracies in the physical parameters \citep{bjoe01}. However, these degeneracies are lifted when Klein-Nishina (KN) effects are included in the SSC cooling calculations \citep[e.g.,][]{lem13,lem15}. 

The signatures of KN effects include harder spectra and shallower light curves at high energies (typically at $\gtrsim1$~keV; \citealt{nas09,jbvdh21}). Such effects are sometimes observed in GRB X-ray light curves. \cite{bndp15} found discrepancies in the energies inferred from X-ray and GeV light curves of 10 GRBs when assuming $\nuc$ is below the X-rays, which can be resolved if IC cooling is KN suppressed in the GeV regime. \citet{lab+18} found a shallower X-ray light curve in GRB~161219B than expected, and attributed the difference due to a change in the evolution of $\nuc$ due to KN effects. The harder-than-expected X-ray spectral indices in the afterglows of GRBs 181201A \citep{lves+19}, 190114C \citep{lag+19}, and 160625B \citep{kfc+20} were all ascribed to KN effects. \citet{kf21} modeled 21 GRBs and found KN effects to be potentially important in at least 4. In the case of SGRB~211106A, \citet{les+22} incorporated IC+KN effects to find much higher values for \epse\ and \dens\ and lower values for \epsb\ than when these effects were ignored. Both shallow X-ray light curves and harder spectra were found for short-duration GRBs (SGRBs) 211106A \citep{les+22} and 210726A \citep{srl+23}, and were shown to be consistent with KN effects in both cases. For SGRBs in particular, IC cooling is predicted \citep{nas09} and observed to be strongly KN-suppressed \citep{flr+21}, further underscoring the importance of incorporating KN corrections in GRB afterglow modeling.

Whereas several efforts have been made to fold in SSC into GRB afterglow modeling, incorporation of KN corrections is rare \citep[e.g.,][]{fd11,bndp15,les+22}, largely due to the complexity of the resulting light curves and spectra and the paucity of simple analytical models that summarize these effects in a way that is easy to implement for data fitting. \citet[][henceforth, NAS09]{nas09} provide a comprehensive reference for KN effects in GRB afterglows. \citet{whl+10} focused on KN corrections to the early high-energy afterglow emission. \citet[][henceforth, JBH21]{jbvdh21} provide analytical approximations for the Compton $Y$-parameter and discuss the impact of KN corrections for a subset of KN-affected regimes. 

In this work, we develop the previous analyses of NAS09 and JBH21 further to delineate new PLSs in the KN regime. We summarize previous work on the $Y$-parameter in the Thompson and KN regimes into a handy reference, including a means of calculating the break frequency ($\nunaught$) where the spectrum converges to the synchrotron-only cooling spectrum. We derive and describe distinct KN spectral regimes and associated PLSs along with their regimes of validity. Finally, we discuss the KN-specific spectral regimes most frequently expected in GRB afterglows and demonstrate how KN corrections help address observed shallow X-ray light curves and hard SEDs. We provide an open-source, self-contained Python package for calculating the KN-corrected $Y$-parameter online, as well as a living, public GitHub repository for the most up-to-date version of the code\footnote{\href{https://github.com/georgeamccarthy/ykn}{https://github.com/georgeamccarthy/ykn} \citep{kn}}. 

This paper is structured as follows. We present the theoretical model in Section 2. We start with the assumptions and approximations used in the model, which we follow with calculations of the Compton $Y$-parameter in the Thomson and KN regimes, expressions for the location of the break frequency where the spectrum returns to synchrotron-only cooling, and expressions for the location of the cooling break. In Section 3, we discuss additional power-law segements in KN-corrected spectra, and provide examples of observational signatures of KN effects in Section 4. We conclude with a summary and plans for future work in Section 5. We use the convention $F_\nu \propto t^\alpha\nu^\beta$ throughout and summarize the symbols used in this work in a glossary in the Appendix. %Table~\ref{tab:glossary}. 
Unless otherwise specified, we assume a flat $\Lambda$CDM cosmology with $\Omega_{M,0}=0.31$,  $\Omega_{\Lambda,0}=0.69$, and $h=0.68$ and use cgs units throughout. 

\section{Klein-Nishina Effects in GRB afterglows}
We begin with a few simplifying approximations. The synchrotron spectrum of a single electron is sharply peaked at the electron's characteristic synchrotron frequency (in cgs units), 
\begin{equation}
    \nusyn(\gammae) \approx \frac{\Gamma \gammae^2 q_e B}{2 \pi m_e c}, 
\end{equation}
where $\Gamma$ is the bulk Lorentz factor of the post-shock fluid, $B$ is the magnetic field strength in the fluid rest frame, $q_e$ is the electron charge $m_e$ is the electron rest mass, and $c$ is the speed of light. 
We follow the framework of \cite{spn98} and assume that electrons are injected behind the shock\footnote{All fluid dynamic quantities used here are measured in the frame of the post-shock fluid unless otherwise specified.} into a power-law distribution in $\gammae$ above a minimum, $\gammam$,
\begin{equation}
    \frac{dn_0}{d\gammae} \propto 
    \begin{cases}
       \gammae^{-p}, & \gammam \le \gammae \le \gamma_{\rm max}  \\
       0, & {\rm otherwise}
    \end{cases}
\end{equation}
Here, $\gamma_{\rm max}$ corresponds to the maximum energy that electrons can attain in the shock acceleration process. For our discussion, however, we ignore this limit, and note that the spectrum is eventually expected to cut off above $\nusyn(\gamma_{\rm max})$. Synchrotron losses will result in the accelerated electrons cooling behind the shock at a rate given by the cooling equation (ignoring IC cooling), 
\begin{equation}
    \frac{d \gammae}{dt^\prime} = -\frac{\sigma_T B^2}{6 \pi m_e c}\gammae^2,
    \label{eq:cooling_rate_no_ic}
\end{equation}
where $t^{\prime} \sim \Gamma t/(1+z)$ is the time in the frame of the post-shock fluid and $t$ is the observer-frame time. 
This yields a characteristic electron Lorentz factor above which electrons are efficiently cooling by synchrotron emission over the lifetime of the system given by,
\begin{equation}
    \gammac = \frac{6\pi m_e c}{\sigma_T B^2 t^\prime}.
\end{equation}
Electron cooling results in an additional break in the electron distribution ($\gammac$) and the associated synchrotron emission spectrum in a way that depends on the relative ordering of $\gammam$ and $\gammac$. In the absence of KN effects and in the regime $\gammam < \gammac$ (slow cooling), we have, 
\begin{equation}
    \frac{dn_0}{d\gammae} \propto 
    \begin{cases}
       \gammae^{-p}, & \gammam \le \gammae \le \gammac  \\
       \gammae^{-p-1}, & \gammac \le \gammae \le \gamma_{\rm max}  \\
       0, & {\rm otherwise},
    \end{cases}
\end{equation}
whereas in the regime 
$\gammac < \gammam$ (fast cooling) we have,
\begin{equation}
    \frac{dn_0}{d\gammae} \propto 
    \begin{cases}
       \gammae^{-2}, & \gammac \le \gammae \le \gammam  \\
       \gammae^{-p-1}, & \gammam \le \gammae \le \gamma_{\rm max}  \\
       0, & {\rm otherwise}.
    \end{cases}
\end{equation}
SSC adds an additional cooling mechanism, which modifies the above electron distributions. This changes the associated synchrotron spectrum, which is the seed photon field for SSC cooling in the first place, making the problem circular. In the presence of SSC, closed-form solutions to the electron distribution are available only in some cases (e.g., for $p=2.5$; \JBH); the general solution (in particular, with the addition of KN effects), requires a series of further approximations. Following previous authors, we now describe the approximations that make this problem tractable and allow us to compute the SSC-corrected synchrotron spectrum in the presence of KN effects, beginning with the SSC cross section in the KN limit. 

From quantum electrodynamics, the $e\gamma\rightarrow e\gamma$ scattering cross section decreases as the energy of the incident photon approaches the electron rest mass in the electron's rest frame (the Klein-Nishina effect). To leading order in $x \equiv h\nu / m_e c^2$, this cross section is given by \citep{rl86}, 
\begin{equation}
    \sigma_{\rm KN} = 
    \begin{cases}
       \sigma_T(1-2x), & x\ll1  \\
       \sigma_T\frac{3(1+2\ln{2x})}{16x}, & x\gg1  \\
    \end{cases}
\end{equation}
where $x$ parameterizes the ratio of incident photon energy to electron rest mass in the frame of the up-scattering electron, and $\sigma_T$ is the Thomson cross section. The KN cross section tends to $\sigma_T$ for low-energy photons (photons that satisfy the limit of $x\ll1$ in the electron rest frame).
We make the step-function approximation for $\sigma_{\rm KN}$, 
\begin{equation}
    \sigma_{\rm KN} = \begin{cases}
        \sigma_T & x < 1 \\
        0        & x \gtrsim 1, 
    \end{cases}
    \label{eq:sigmaknstep}
\end{equation}
which is equivalent to assuming that for photons with energies greater than the electron rest mass in the frame of the up-scattering electron, IC scattering is negligible\footnote{This approximation breaks down when the energy density of the seed photon field (here, the synchrotron spectrum) has a steeply rising spectrum, $d\ln{F_\nu}/d\ln\nu>1$ (\NAS), such as for $\nu<\nusa$ in the synchrotron self-absorbed (SSA) regime. We note that SSC cooling is a local effect, since electrons behind the shock upscatter the local photon field, whereas SSA is a global effect due to the optical depth of the plasma. A complete calculation requires solving the radiative transfer equation using the local photon field (including KN effects) at every point along each light ray, which is beyond the scope of the present work.}.

Following, \NAS, the effect of this approximation can be encapsulated in the following function of an electron's Lorentz factor,
\begin{equation}
    \gammaehat = \frac{m_e c^2 \Gamma}{h \nusyn(\gammae)} \propto \gammae^{-2},
\end{equation}
and its reciprocal function
\begin{equation}
    \gammaetilde = \left( \frac{\gammae m_e c^2 \Gamma}{h \nusyn(\gammae)}\right)^{1/2},
\end{equation}
where $h$ is Planck's constant. Electrons with Lorentz factor $\gammae$ emit $\nusyn(\gammae)$ photons that can be up-scattered by electrons with energies up to $\gammaehat$ and, themselves can up-scatter emission by electrons of Lorentz factors up to $\gammaetilde$. This introduces several new critical Lorentz factors in addition to $\gammam$ and $\gammac$, including,
\begin{equation}
    \gammamhat = \frac{m_e c^2 \Gamma}{h \nu_m},
\end{equation}
the maximum-energy electrons capable of scattering $\numax$ photons,
\begin{equation}
    \gammachat = \frac{m_e c^2 \Gamma}{h \nu_c},
\end{equation}
the maximum-energy electrons capable of scattering $\nuc$ photons, and
\begin{equation}
    \gammaself = \left(\frac{B_{\rm QED}}{B}\right)^{1/3} = \gammae^{2/3}\gammaehat^{1/3},
\end{equation}
where $B_{\rm QED} = 2\pi m_e^{2} c^3 / (q_e h) = 4.41 \times 10^{13}$ G is the quantum critical field; these electrons up-scatter their own emission, such that $\hat{\gamma}_{\rm self} = \tilde{\gamma}_{\rm self} = \gammaself$. Each of these critical Lorentz factors may result in additional breaks in the observed synchrotron spectrum at the corresponding value of $\nusyn(\gammae)$ depending on their relative ordering with respect to $\gammac$ and the value of the Compton $Y$-parameter at each break, which we discuss next. 

\subsection{The Compton Y-parameter}
The importance of SSC cooling is given by the Compton $Y$-parameter, which is defined as the ratio of the SSC to synchrotron power, 
\begin{equation}
    Y(\gammae) \equiv \frac{P_{\rm SSC}(\gammae)}{P_{\rm syn}(\gammae)},
\label{eq:y_def}
\end{equation}
and is, in general, a function of the electron Lorentz factor, $\gammae$. SSC effects modify the electron radiative cooling equation to,
\begin{equation}
    \frac{d \gammae}{dt^\prime} = \frac{\sigma_T B^2}{6 \pi m_e c}\gammae^2 [1 + Y(\gammae)].
\label{eq:cooling_rate_with_ic}
\end{equation}
IC cooling significantly affects the cooling rate only if $Y(\gammae) \gg 1$. If $Y\ll1$ for all $\gammae$, then IC-cooling can be ignored entirely.
We discuss the form of $Y$ separately in the regime where IC cooling is dominant (the Thomson regime; Section~\ref{text:YT}) and where IC cooling is KN-suppressed (Section~\ref{text:YKN}).  

\subsubsection{Compton $Y$ in the Thomson regime, $\YT$}
\label{text:YT}
In the presence of IC cooling and the absence of KN effects, $Y(\gammae)=\YT$ (the Compton $Y$-parameter in the Thomson limit), which is independent of $\gammae$. Knowledge of \YT\ is also useful in the KN-suppressed regime since it sets the maximum value $Y(\gammae)$ can achieve at any given time. \citet[][hencefoth, SE01]{se01} derived an expression for \YT\ starting with an approximate form of equation~\ref{eq:y_def}, but ignoring the dependence of \YT\ on the shape of the electron distribution. \JBH\ derived a more accurate expression for \YT\ by integrating over $\langle\gammae^2\rangle$, which accounts for the fact that the mean ratio of the inverse Compton to synchrotron luminosity depends on the shape of the electron distribution. 

\begin{figure}
    \centering
    \includegraphics[width=\columnwidth]{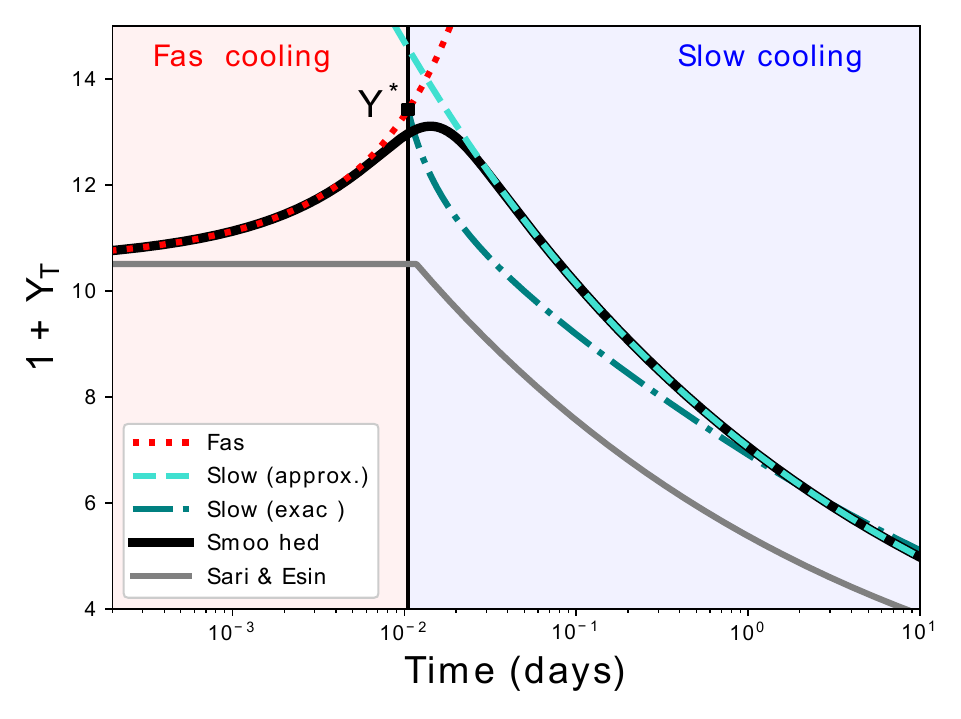}
    \caption{The Compton $Y$-parameter in the Thomson regime (\YT) as a function of time spanning before and after the fast-to-slow cooling transition (vertical line) using the formalism of \JBH\ (black), calculated by smoothly connecting a closed-form solution for \YT\ in fast cooling (red, dotted) with an asymptotic closed-form approximation for \YT\ in the slow cooling regime (blue, dashed). For comparison, we also show the exact solution for \YT\ in slow cooling (teal, dash-dot) and \YT\ derived using the prescription of SE01 (grey). The offset between \JBH\ and SE01 is due to differences in the definition of $Y$. The parameters adopted for this calculation are: $p=2.5$, $z=1$, $\epsilone=0.1$, $\epsilonb=10^{-3}$, $E_{\rm K,iso}=10^{53}$~erg, and $n_0=5$ in an ISM environment.}
    \label{fig:ytplot}
\end{figure}

We plot \JBH's solution and its constituent components in Figure~\ref{fig:ytplot}, where we also compare it to the canonical $\YT$ derived by SE01. We find that the asymptotic value of \YT\ from \JBH\ in the deep slow cooling regime ($\gammam\ll\gammac$) is larger than that of SE01 by a factor of $(3-p)^{-1/(4-p)}$ ($\approx1.7$ for $p=2.5$) for $\YT\gg1$, and a factor of $1/(3-p)$ for $\YT\ll1$. This offset can be traced to the definition of $Y$ discussed above. 
Whereas the expression derived by \JBH\ is more accurate for $2<p<3$, SE01's solution is valid for $p>3$. In this work, we use the \YT\ expressions derived by \JBH. The details of our implementation are described in Appendix~\ref{appendix:YT} and the corresponding equations are given in Table~\ref{table:Yt} in that Appendix. 

\subsubsection{Compton $Y$ in the Klein-Nishina Regime, $Y(\gammae)$}
\label{text:YKN}
Electrons with $\gammae \ll \min{(\gammamhat, \gammachat)}$ are able to efficiently upscatter the majority of the photon field and thus $Y(\gammae)=\YT$ for SSC cooling in the Thomson regime. When $\nusyn^{\rm peak} < \min{(\numhat, \nuchat)}$, $Y_{\rm T, KN}$ will not deviate significantly from $Y_{\rm T, no-KN}$. In other words, the inclusion of KN breaks in the electron spectrum does not significantly alter $\YT$ in the KN regime\footnote{In the case where $\nusyn^{\rm peak} > \min{(\numhat, \nuchat)}$, additional KN breaks (e.g., $\hat{\nu}_0$ and $\hat{\hat{\nu}}_{\rm m}$) must be considered (NAS09). We do not discuss these cases further in the present work.}. When KN corrections are important, the $Y$-parameter decreases with $\gammae$ from its value of $Y(\gammae)\approx\YT$ at $\gammae < \min{(\gammamhat,\gammachat)}$ to $Y\ll1$ at $\gammae \gg \max{(\gammamhat,\gammachat)}$. If $\YT<1$, then SSC cooling can be ignored entirely and the spectrum can be well-approximated by the standard synchrotron spectrum in the absence of IC cooling (henceforth, the synchrotron-only cooling spectrum, SOCS). If $\YT>1$ on the other hand, then $Y$ crosses unity at a well-defined value of $\gammae=\gammanaught$ (Section~\ref{text:gammanaught}) and the spectrum returns to the SOCS at $\nu>\nunaught$. Below $\nunaught$, the spectrum transitions from SSC-dominated cooling at $\nuc<\nu\ll\nunaught$, where $Y\approx\YT$, to synchrotron-only cooling at $\nu\gtrsim\nunaught$, through a series of additional PLSs (\NAS). In our implementation, when $\YT<1$ (equivalent to $\nunaught<\nuc$) we ignore SSC and employ the standard synchrotron spectrum \citep{gs02}. In the remainder of this section, we discuss the calculation of $Y(\gammae)$ for the scenario $\nuc<\nunaught$, where IC and KN effects are both important.

Calculations of the KN-corrected Y-parameter for $\gammac<\gammanaught$ are broken down into two key regimes, the weak KN regime ($\gammam < \gammamhat$) and the strong KN regime ($\gammamhat < \gammam$).
In the following, we focus our attention on the weak-KN regime\footnote{The definition of weak and strong KN regimes varies between the two key publications referenced within this work (\NAS\ and \JBH). We use the definition in \NAS. We note that \JBH\ also focus on the weak-KN regime ($\gammam<\gammamhat$) and do not discuss the $\gammamhat<\gammam$ case.}. We note that in the slow-cooling, strong-KN case, SSC cooling does not affect the electron distribution and the spectrum returns to SOCS (\NAS), and defer a description of KN-corrected spectra in the strong-KN, fast-cooling regime to future work.

Under the step-function approximation for $\sigma_{\rm KN}$ (Equation~\ref{eq:sigmaknstep}), $Y(\gammae)$ is found by convolving the synchrotron emission function of a single electron $P_\nu$ with the electron distribution up to the maximum scattering energy $\tilde{\nu}(\gammae)$ to yield (\JBH, \NAS),
\begin{equation}
    Y(\gammae) \propto \int_0^{\tilde{\nu}^\prime(\gammae)} d \nu^\prime \int d\gammastar P_{\nu^\prime}(\gammastar) \frac{d n_0}{d \gammastar}.
\label{eq:full_y_equation}
\end{equation}
Here $\tilde{\nu}^{\prime}(\gammae)\equiv\nusyn(\tilde{\gamma})\propto\tilde{\gamma}^2\propto\gammae^{-1}$ is measured in the fluid rest frame. Following \JBH, we approximate the single electron emissivity as,
\begin{equation}
    P_{\rm \nu}(\gammastar) \propto
    \begin{cases}
        \delta(\nu - \nusyn(\gammastar))\nusyn(\gammastar), & \nu \gtrsim \nusyn(\gammastar), \\
        \nu^{1/3}, & \nu \ll \nusyn(\gammastar),
    \end{cases}
    \label{eq:single_electron_emissivity}
\end{equation}
When $\tilde{\nu}^\prime(\gammae)<\min{(\nuc,\numax)}$, which corresponds to $\nusyn(\gammae)>\max{(\numhat,\nuchat)}$, this yields,
\begin{equation}
    Y(\gammae) \propto \int_0^{\tilde{\nu}^\prime(\gammae)} d\nu^\prime {\nu^\prime}^{\frac13} \int d\gammastar  \frac{d n_0}{d \gammastar} \propto \left[\tilde{\nu}^\prime(\gammae)\right]^{\frac43}\propto\gammae^{-\frac43}.
\end{equation}
Conversely, following \JBH, when $\tilde{\nu}^\prime(\gammae)>\min{(\nuc,\numax)}$, which corresponds to $\nusyn(\gammae)<\min{(\numhat,\nuchat)}$, we set $Y=\YT$. 
Otherwise, from the properties of the Dirac delta, we have,
\begin{equation}
    Y(\gammae) \propto \int_1^{\gammaetilde} d\gammae \gammae^2 \frac{d n_0} {d \gammae}.
    \label{eq:y_proportionality}
\end{equation}
While this is a more complex expression,  power-law segments can be found for $Y$ in asymptotic regimes. Because $\gammaehat \propto \gammae^{-2}$, we will always have $\gammamhat<\gammachat$ in fast cooling and $\gammachat<\gammamhat$ in slow cooling. In either cooling regime, if $\gammaetilde < {\rm min}\{\gammac, \gammam\}$, the $\nu^\frac13$ tail of the electrons' emission dominates and we have already found $Y(\gammae) \propto \gammae^{-\frac43}$. The orderings with ${\rm max}\{\gammac, \gammam\} < \gammaetilde$, are the Thomson-KN transitional regimes where for $\gammae \ll {\rm min}\{\gammachat, \gammamhat\}$, $Y(\gammae) = \YT$ (see \JBH\ for further discussion). We always approximate $Y(\gammae) = \YT$ in this regime. The remaining regime to consider is ${\rm min}\{\gammac, \gammam\} < \gammaetilde < {\rm max}\{\gammac, \gammam\}$, which we discuss next.

In fast cooling, the electron distribution accounting for SSC cooling can be approximated by (\NAS),
\begin{equation}
\label{eq:e_dist_fast}
\frac{dn_0}{d\gammae} \propto \frac{1}{1+Y(\gammae)}
\begin{cases}
    \gammae^{-2}, &\gammac < \gammae < \gammam\\
    \gammae^{-p-1}, &\gammam < \gammae\\
\end{cases}
\end{equation}
For the case $\gammac < \gammaetilde < \gammam$ (equivalently $\gammamhat < \gammae < \gammachat$) we integrate,
\begin{equation}
    Y(\gammae) \propto \int_{\gammac}^{\gammaetilde} d\gammae \gammae^2 \gammae^{-2}. 
\end{equation}
For $\gammac \ll \gammam$, this yields $Y(\gammamhat < \gammae < \gammachat) \propto \gammaetilde \equiv \gammae^{-\frac{1}{2}}$ (\JBH).
In slow cooling,
the electron distribution can be approximated by,
\begin{equation}
\label{eq:e_dist_slow}
\frac{dn_0}{d\gammae} \propto
\begin{cases}
    \gammae^{-p}, &\gammam < \gammae < \gammac\\
    \frac{1}{1+Y(\gammae)}\gammae^{-p-1}, &\gammac < \gammae,\\
\end{cases}
\end{equation}
where IC only affects cooling electrons ($\gammac < \gammae$). 
For $\gammam < \gammaetilde < \gammac$ (equivalently $\gammachat < \gammae < \gammamhat$) we integrate,
\begin{equation}
    Y(\gammae) \propto \int_{\gammac}^{\gammaetilde} d\gammae \gammae^2 \gammae^{-p}. 
\end{equation}
For $\gammam \ll \gammac$, this yields $Y(\gammamhat < \gammae < \gammachat) \propto \gammaetilde^{3-p} \equiv \gammae^{(p-3)/2}$ (\JBH).

In summary, we have in the weak KN regime ($\gammam<\gammamhat$),
\begin{equation}
    Y(\gammae)= Y(\gammamhat)
    \begin{cases}
    1, & \gammae < \gammamhat < \gammachat\\
    \left(\frac{\gammae}{\gammamhat}\right)^{-\frac12}, & \gammamhat < \gammae < \gammachat\\   
    \left(\frac{\gammachat}{\gammamhat}\right)^{-\frac12}\left(\frac{\gammae}{\gammachat}\right)^{-\frac43}, & \gammamhat<\gammachat<\gammae,
\end{cases}
\label{eq:YKN-weak-fast}
\end{equation}
in fast cooling, and 
\begin{equation}
    Y(\gammae)= Y(\gammachat)
    \begin{cases}
    1, & \gammae < \gammachat < \gammamhat\\
    \left(\frac{\gammae}{\gammachat}\right)^{\frac{p-3}{2}}, & \gammachat < \gammae < \gammamhat\\   
    \left(\frac{\gammamhat}{\gammachat}\right)^{\frac{p-3}{2}}\left(\frac{\gammae}{\gammamhat}\right)^{-\frac43}, & \gammachat<\gammamhat<\gammae,
\end{cases}
\label{eq:YKN-weak-slow}
\end{equation}
in slow cooling, where we approximate $Y(\gammamhat)\approx\YT$ and $Y(\gammachat)\approx\YT$ in both cases. 

\label{text:gammanaught} 
Since $Y(\gammae)\le\YT$, if $\YT>1$, then KN effects will return $Y(\gammae)$ to unity at a well-defined $\gammanaught$, such that
\begin{equation}
    Y(\gammanaught) = 1. 
\end{equation}
We use the expressions for $Y$ in Equations~\ref{eq:YKN-weak-fast} and \ref{eq:YKN-weak-slow} to solve for $\gammanaught$, and find,
\begin{align}
\gamma_0 &= 
\begin{cases}
    \YT^2\gammamhat & \gammamhat < \gamma_0 < \gammachat\\
    \left(\YT\frac{\gammac}{\gammam}\right)^{3/4}\gammachat &\gammachat < \gamma_0,
\end{cases}
\end{align}
in fast cooling, and 
\begin{align}
\gamma_0 &= 
\begin{cases}
    \YT^{\frac{2}{3-p}}\gammachat, & \gammachat < \gamma_0 < \gammamhat\\
    \YT^\frac{3}{4}\gammamhat\left(\frac{\gammamhat}{\gammachat}\right)^\frac{3(p-3)}{8}, &\gammamhat < \gamma_0,
\end{cases}
\end{align}
in slow cooling. These expressions yield $\nunaught=\nusyn(\gammanaught)$, above which the spectrum returns to the SOCS. 

\subsection{\Yc\ and the location of the cooling break}
\begin{table*}[ht]
\caption{$Y_c$ in the weak KN regime ($\gammam < \gammamhat$).} 
\centering 
\begin{tabular}{c c l} 
\hline\hline 
Rule 1 & Rule 2 & $Y_c$\\ [2ex]
\hline
$\gammac<\gammam$ & & \YT\ \\
$\gammam<\gammac$ & $\gammac<\gammachat<\gammamhat$ & \YT\ \\
$\gammam\ll\gammac$ & $\gammachat<\gammac<\gammamhat$ & 
$\left[\frac{\epse}{\epsb}\left(\frac{p-2}{3-p}\right)\left(\frac{\gammam}{\gammacs}\right)^{p-2} \left(\frac{\gammacs}{\gammacshat}\right)^{\frac{p-3}{2}}\right]^{\frac{2}{p-1}}
(\Yc \gg 1)$ \\
& & $\frac{\epse}{\epsb}\left(\frac{p-2}{3-p}\right)\left(\frac{\gammam}{\gammacs}\right)^{p-2} \left(\frac{\gammacs}{\gammacshat}\right)^{\frac{p-3}{2}}
(\Yc \ll 1)$ \\
$\gammam\ll\gammac$ & $\gammachat<\gammamhat<\gammac$ & $< 1$ \\
\hline
\end{tabular}
\label{table:Yc} 
\end{table*}
SSC increases the cooling rate of $\gammae$ electrons by a factor $1 + Y(\gammae)$ (equation \ref{eq:cooling_rate_with_ic}). Therefore, electrons cooling through SSC emission over the lifetime of the system cool to energies that are a factor $(1 + \Yc)$ lower than they would have cooled to through synchrotron emission alone ($\gammacs$),
\begin{equation}
    \gammac = \frac{\gammacs}{1 + \Yc},
\end{equation}
where $\gammacs$ is the critical cooling Lorentz factor in the case of no IC cooling, and $\Yc \equiv Y(\gammac)$. Note that $\nu_c \equiv \nusyn(\gamma_c)$ depends on \Yc.
We follow the general steps laid out in \JBH\ to compute \Yc\ in the weak KN regime ($\gammam<\gammamhat$), although some of our results are different from theirs. In the fast-cooling, weak-KN regime, we have $\gammac<\gammam<\gammamhat<\gammachat$, for which we approximate $\Yc \approx \YT$ (Section~\ref{text:YKN}). 
In the slow-cooling, weak-KN regime, there are three possible locations for $\gammac$ with respect to the $\gammamhat$ and $\gammachat$ breaks that preserve the order, $\gammachat<\gammamhat$. These are,  $\gammac<\gammachat<\gammamhat$, $\gammachat<\gammac<\gammamhat$, and $\gammachat<\gammamhat<\gammac$. In the first case, we again approximate $\Yc\approx\YT$ (Section~\ref{text:YKN}). In the last case, we will always find $\Yc < 1$ (\NAS), and the spectrum returns to the SOCS. In the case $\gammachat<\gammac<\gammamhat$, we have from equation \ref{eq:YKN-weak-slow}, 
\begin{equation}
    \Yc = Y(\gammachat) \left(\frac{\gammac}{\gammachat}\right)^{\frac{p-3}{2}}.
\label{eq:Yc-slow-JBHweak}
\end{equation}
To derive $Y(\gammachat)$ in this regime, we must first calculate $Y(\gammae)$ by integrating over the electron distribution. The result is given in equation B12 of \JBH, 
\begin{align}
Y(\gammae)&=\frac{\epse}{\epsb(3-p)(1+\Yc)}\left(\frac{\gammam}{\gammac}\right)^{p-2}\nonumber\\
&\times\left[1-\frac{p-2}{3-p}\left(\frac\gammam\gammac\right)^{3-p}+(p-3)\left(\frac\gammae\gammachat\right)^{\frac{p-2}{2}}\right]\nonumber\\
&\times\left[1-\frac1p\left(\frac\gammam\gammac\right)^{p-1}\right]^{-1}, 
\end{align}
from which we have, 
\begin{align}
Y(\gammachat)&=\frac{\epse(p-2)}{\epsb(3-p)(1+\Yc)}\left(\frac{\gammam}{\gammac}\right)^{p-2} \nonumber \\
&=\frac{\epse(p-2)(1+\Yc)^{p-3}}{\epsb(3-p)}\left(\frac{\gammam}{\gammacs}\right)^{p-2} 
\end{align}
at $\gammae=\gammachat$ and under the ultra-slow-cooling ($\gammam\ll\gammac$) approximation. Combining this with Equation~\ref{eq:Yc-slow-JBHweak} and substituting $\gammac = \gammacs (1 + \Yc)^{-1}$ and $\gammachat = \gammacshat(1 + \Yc)^2$, we find,
\begin{equation}
    \Yc(1 + \Yc)^{\frac{(p-3)}{2}} = \frac\epse\epsb\left(\frac{p-2}{3-p}\right)\left(\frac{\gammam}{\gammacs}\right)^{p-2}\left(\frac{\gammacs}{\gammacshat}\right)^{\frac{p-3}{2}}.
\end{equation}
This yields, 
\begin{align}
\Yc &= \begin{cases}
    \left[\frac\epse\epsb\left(\frac{p-2}{3-p}\right)\left(\frac{\gammam}{\gammacs}\right)^{p-2}\left(\frac{\gammacs}{\gammacshat}\right)^{\frac{p-3}{2}}\right]^{\frac{2}{p-1}}, & \Yc\gg1\\
   \frac\epse\epsb\left(\frac{p-2}{3-p}\right)\left(\frac{\gammam}{\gammacs}\right)^{p-2}\left(\frac{\gammacs}{\gammacshat}\right)^{\frac{p-3}{2}}, &\Yc\ll1,
\end{cases}
\end{align}
which is similar to (but slightly different from) the expressions derived by \JBH\ in their Table 4. 
Our results for $\Yc$ are summarized in Table \ref{table:Yc}.

\section{Klein-Nishina corrected spectra}
\begin{figure*}[htbp]
    \centering
    \begin{tabular}{cc}
        \centering
        \includegraphics[width=0.44\textwidth]{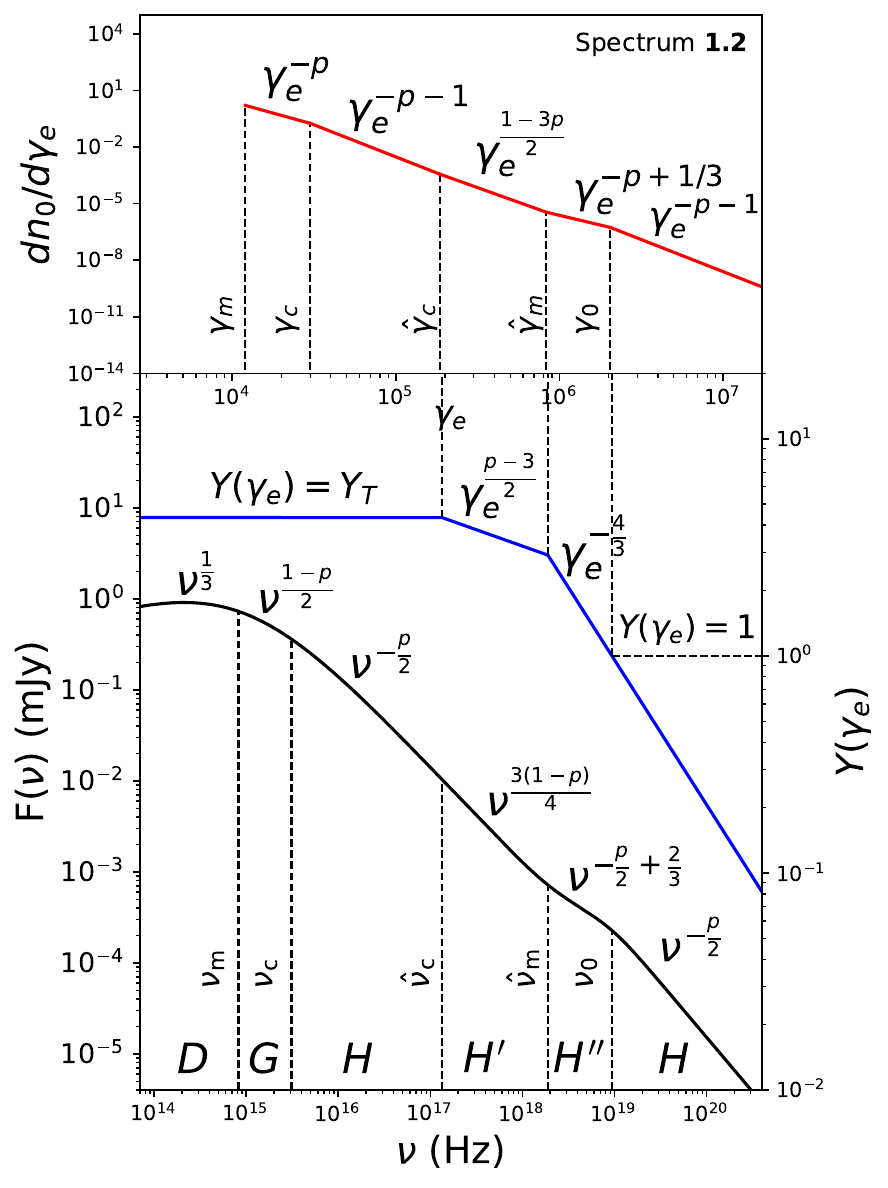} &
        \includegraphics[width=0.44\linewidth]{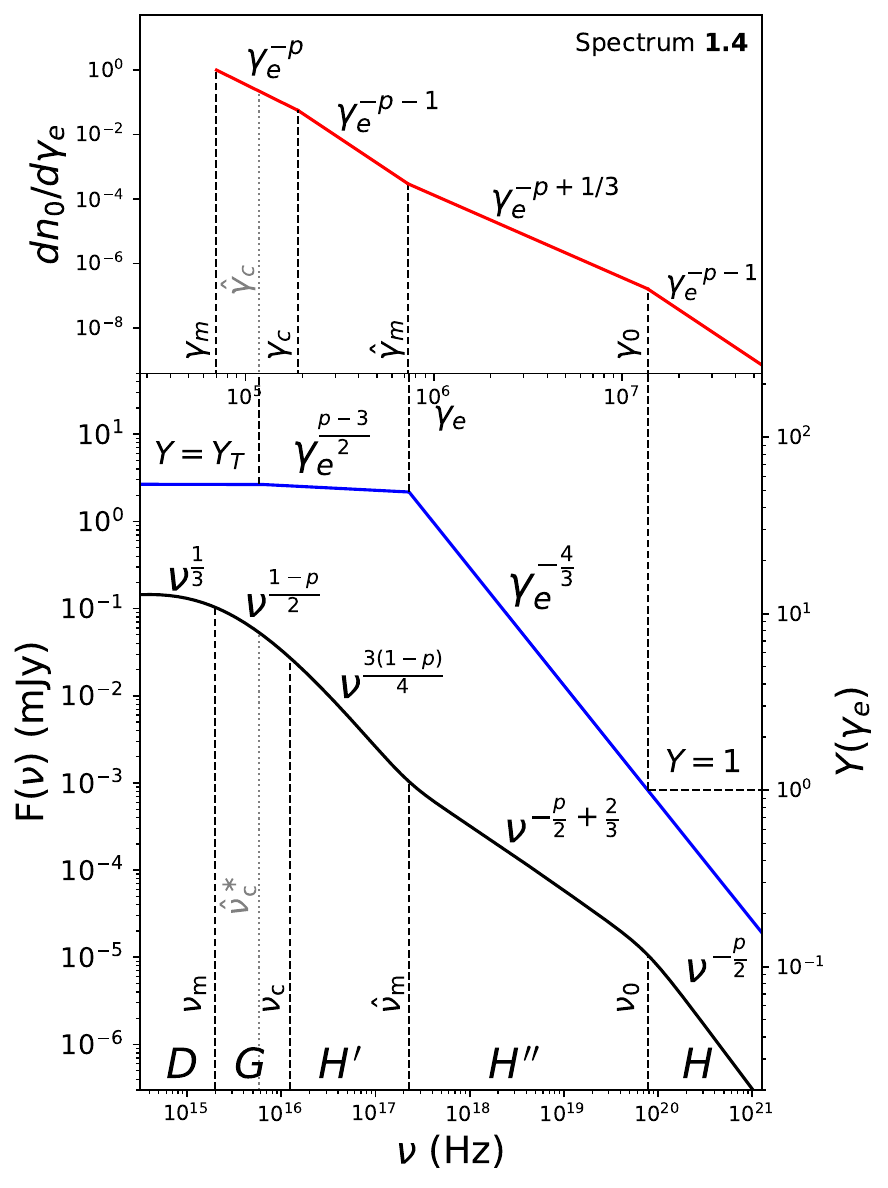}\\
        \includegraphics[width=0.44\linewidth]{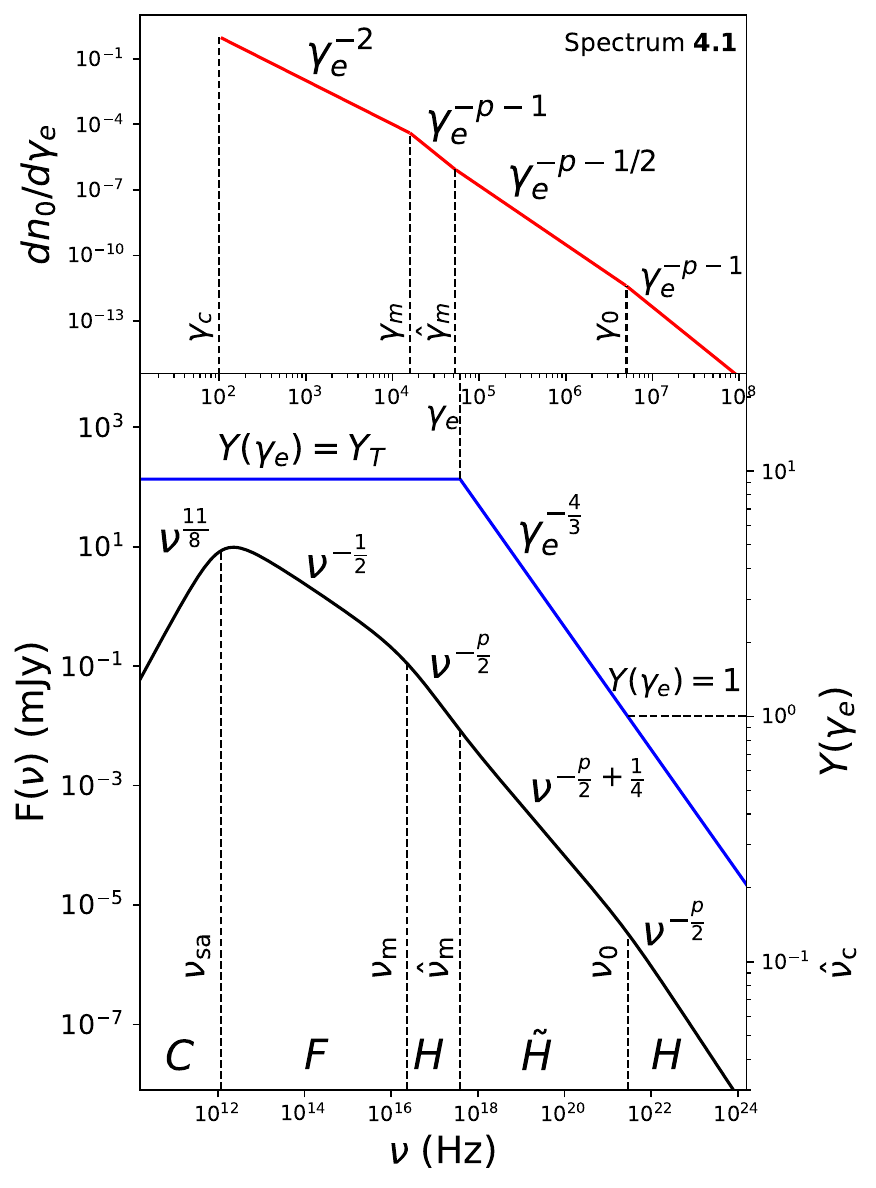}&
        \includegraphics[width=0.44\linewidth]{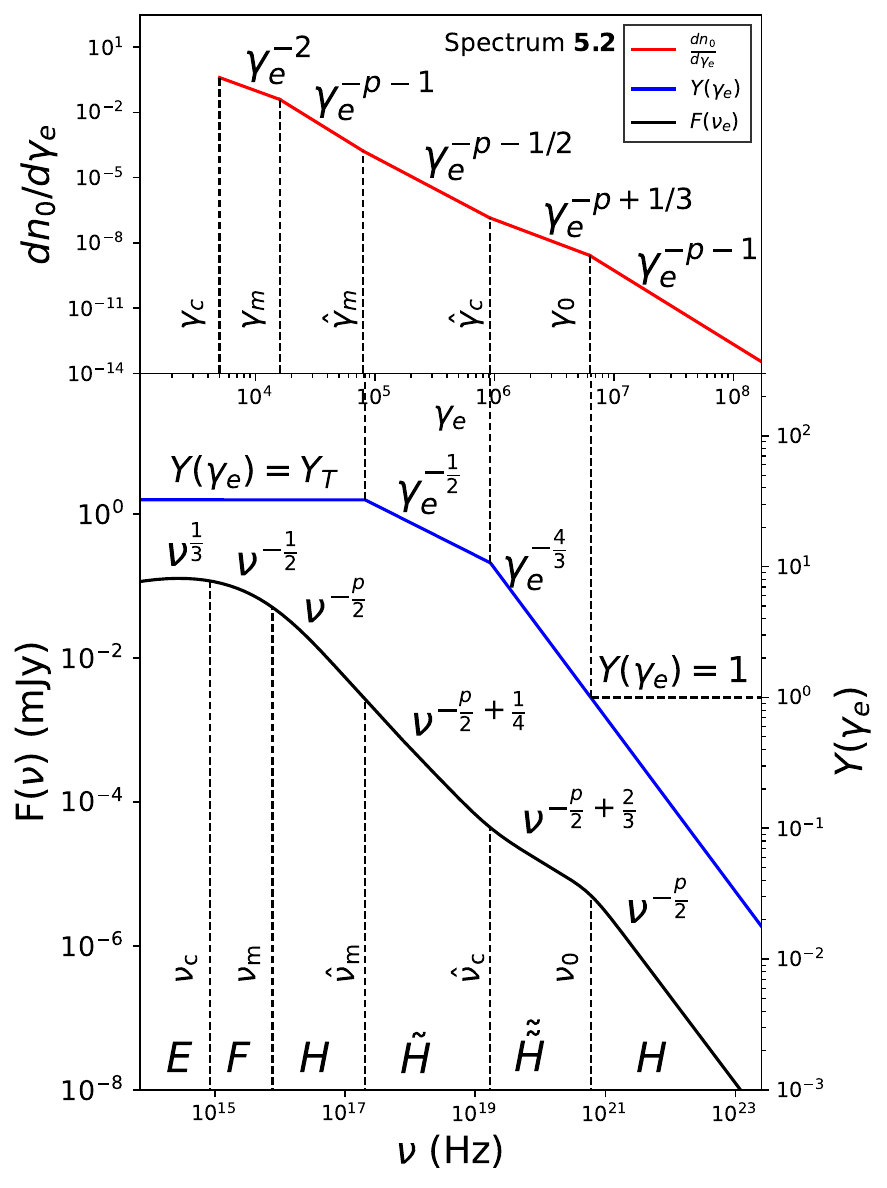}
    \end{tabular}
    \caption{Example electron distributions (${dn_0}/{d \gammae}$; top panels, red power laws, normalized to 1 at the lowest Lorentz factor) with corresponding Compton $Y$ ($Y(\gammae)$; bottom panels, blue power laws), and KN-corrected synchrotron spectrum ($F(\nu$); bottom panels, black curves) in slow cooling (top row) and fast cooling (bottom row). We mark all critical Lorentz factors for ${dn_0}/{d \gammae}$ and $Y(\gammae)$ (the latter being $\gammamhat$, $\gammachat$, and $\gammanaught$) in each subplot.  Breaks in $Y(\gammae)$ correspond to changes in the SSC photon field ($\tilde{\nu}_e$ passing below $\{\numax, \nuc\}$), yielding corresponding breaks in $F(\nu)$ at $\{\numhat, \nuchat\}$, which are marked in the lower panels when visible. At $\nunaught$ (where $Y(\gammanaught) = 1$), the spectrum returns to the SOCS. The parameters used in each subplot are listed in the caption to Figure~\ref{fig:kn-nokn-comparison}.}
    \label{fig:kn_diagram}
\end{figure*}

The observed synchrotron flux density $F(\nu)$, at frequencies where the source is optically thin ($\nusa < \nu$), is proportional to the radiated synchrotron power per unit volume per unit frequency in the local rest frame of the fluid, $P_\nu$,
\begin{equation}
    F_\nu \propto P_\nu = \int_{\gamma_{\rm min}}^{\gamma_{\rm max}} P_\nu(\gammae) \frac{d n_0}{d \gammae} d\gammae.
\label{eq:synchrotron_power}
\end{equation}
The power-law electron distributions (equations \ref{eq:e_dist_fast} and \ref{eq:e_dist_slow}) result in spectra comprising several PLSs smoothly joined at the break frequencies, with the normalization set by the $F_{\nu}(\numax)$. We derive KN corrections to the 5 spectral shapes (1 -- 5) described by \citet[][GS02]{gs02}, who present a full description of how these spectra are calculated without KN-corrected SSC. To compute KN-corrected spectra, we need to know the locations of the KN break frequencies ($\numhat$, $\nuchat$, $\nunaught$), and the spectral index on each spectral segment.
We can calculate the spectral index, $\beta(\gammae)$,
\begin{equation}
    \beta(\gammae) = \frac{d ln(F_\nu)}{d ln(\nu)}\bigg|_{\nu_{\rm syn}(\gammae)}
\end{equation}
from the KN-corrected electron Lorentz factor distribution $d n_0 / d \gammae$ (equations~\ref{eq:e_dist_fast} and \ref{eq:e_dist_slow}),
\begin{equation}
    \beta(\gammae) = \frac{1}{2} \left( \frac{d\text{ln}\left(\frac{d n_0}{d \gammae} \right)}{d\text{ln}(\gammae)} + 1 \right).
\end{equation}
The KN Y-parameter derived in section \ref{text:YKN}, introduces additional PLSs in $d n_0 / d \gammae$. In fast cooling, 
\begin{equation}
    \frac{d n_0}{d \gammae} \propto
    \begin{cases}
    \gammae^{-p-\frac12}, & \gammamhat<\gammae<\gammachat, \\
    \gammae^{-p+\frac13}, & \gammamhat<\gammachat<\gammae, \\
\end{cases}
\end{equation}
If $\Yc$ and $\nunaught$ are sufficiently large, the ordering $\numhat < \nu < \nuchat$ results in a spectral index $\beta = -\frac{p}{2} + \frac{1}{4}$ which we label the $\tilde{H}$ segment, and $\numhat < \nuchat < \nu$ results in spectral index $\beta = -\frac{p}{2} + \frac{2}{3}$ which we label $\tilde{\tilde{H}}$. In slow-cooling,
\begin{equation}
    \frac{d n_0}{d \gammae} \propto
    \begin{cases}
    \gammae^{(1-3p)/2}, & \gammachat<\gammae<\gammamhat, \\
    \gammae^{-p+\frac13}, &\gammachat<\gammamhat<\gammae, \\
\end{cases}
\end{equation}
The ordering $\nuchat < \nu < \numhat$ results in $\beta = -\frac{3}{4}(p-1)$, which we refer to as $\Hprime$, and $\numhat < \nu$ results in $\beta = -\frac{p}{2} + \frac{2}{3}$ which we label $\Hprimeprime$. 
These KN-corrected, SSC-dominated spectral segments and corresponding relevant spectral orderings are shown in Table \ref{table:weak_kn_spectral_segments}. We present diagrams of $Y(\gammae)$ and $dn_0/d\gamma_e$, along with corresponding KN-corrected synchrotron spectra (described in the next section) for some typical cases in Figure~\ref{fig:kn_diagram}. 

\begin{table}[ht]
\caption{Spectral segments}
\centering
\begin{tabular}{c l c c}
\hline\hline
Cooling & Spectral & Spectral & Spectral \\
Regime & Segment & Index, $\beta$ & Ordering \\ [2ex]
\hline
\multicolumn{4}{c}{Non-KN Breaks}\\
* & A & 5/2 & $\{\numax,\nuc\} < \nusa$ \\
* & B & 2   & all \\
Fast & C & 11/8     & $\{\nuc,\nusa\}<\numax$ \\
Slow & D & 1/3      & $\nusa<\numax<\nuc$ \\
Fast & E & 1/3      & $\nusa<\nuc<\numax$\\
Fast & F & -1/2     & $\{\nuc,\nusa\}<\numax$ \\
Slow & G & (1-p)/2  & $\{\nusa,\numax\}<\nuc$ \\
\hline
\multicolumn{4}{c}{KN Breaks}\\
Slow & $H$ & $-p/2$ & $\nuc < \nu < \nuchat$ \\
    &&& or \{\nuc, \nunaught\} $< \nu$\\
 & $H^{\prime}$ & $3(1-p)/4$ & $\{\nuc,\nuchat\} < \nu < \{\numhat,\nunaught\}$ \\
& $H^{\prime \prime}$ & $-p/2 + 2/3$ & $\{\nuc,\numhat\} < \nu < \nunaught$\\ 
Fast &$H$ & $-p/2$ & $\nuc < \nu < \numhat$ \\
    &&& or \{\nuc, \nunaught\} $< \nu$\\
&$\tilde{H}$ & $-p/2 +1/4$ & $\{\nuc,\numhat\} < \nu < \{\nuchat,\nunaught\}$ \\
&$\tilde{\tilde{H}}$ & $-p/2 + 2/3$ & $\{\nuc,\nuchat\} < \nu < \nunaught$\\ 
\hline
\end{tabular}
\tablecomments{Break frequencies in braces can appear in any order. Spectral segment B is always present. Segments A and B can appear in either fast or slow cooling spectra.}
\label{table:weak_kn_spectral_segments}
\end{table}

\subsection{Derivation of the distinct KN spectra}
\label{text:derivation_of_kn_spectra}
The presence of KN breaks greatly increases the number of possible spectral shapes. Fortunately, many orderings of \nuc, \numax, \nuchat, \numhat, and \nunaught\ produce spectra that are not distinct from each other or from the SOCS. KN-corrected spectra only differ significantly from the SOCS if $Y(\gammac) \gg 1$. In this case, changes to the spectrum are visible\footnote{KN corrections only have observable effects on the synchrotron spectrum at $\nuc<\nu$. Thus, any KN break frequency (e.g. $\nuchat$ and $\numhat$) that falls redward of $\nuc$ is invisible.} between $\nuc$ and $\nunaught$. We compile a list of the orderings that do result in distinct spectra from the SOCS and describe their spectral shapes in this section. We summarize our results in Table \ref{table:kn_spectra}. Spectral breaks labeled with an asterisk (e.g. $\nuchat^\ast)$ are not visible breaks in the spectrum, however, their relative ordering to the other breaks is important. For example, in spectrum 1.1, $\numhat^\ast$ is not an observable break because $\nunaught < \numhat$ and the KN spectrum returns to the synchrotron spectrum above $\nunaught$. Two breaks appear in brackets when the relative ordering of those two breaks results in the same spectral shape. For example, in spectrum 1.3, KN effects only affect the spectrum above $\nuc$; therefore, the relative ordering of $\{\numax,\nuchat\}<\nuc$ will not change the ordering of spectral segments, which is: $G, H', H$, so long as $\numax$ and $\nuchat$ are both below $\nuc$ in either order ($\nuchat^*$ will be invisible and the break will occur at $\numax$ in both cases).

In slow cooling, $\numax < \nuc$ and $\nuchat<\numhat$ (Section~\ref{text:YKN}), and $\Yc \gg 1$ requires $\nuc < \nunaught$.
Additionally, if $\nuc < \numhat$ then $\Yc < 1$ (\NAS).  The $\nunaught$ and $\numhat$ breaks may occur in either order, therefore only the orderings with $\{\nu_c,\nuchat\} < \{\nu_0,\numhat\}$ have an observable signature. For spectral shape 1, which has $\nusa < \numax$, swapping $\{\nu_c,\nuchat\}$ and $\{\nu_0,\numhat\}$ results in four possible spectral orderings (1.1 -- 1.4). Spectral shape 2 has $\numax < \nusa$, the KN corrections only affect the spectrum above $\nuc$. Therefore, spectra 2.1--2.4 are the same as 1.1--1.4 except that $\numax$ is exchanged for $\nusa$ and segment D for segment A.

In fast-cooling, $\nuc < \numax$ and $\numhat < \nuchat$.
The weak KN regime, therefore, only allows four orderings that deviate from the SOCS: $\{\nusa,\nuc\}<\numax<\numhat<\{\nunaught,\nuchat\}$. These are spectra 4.1, 4.2 in the case of $\nuc < \nusa$ and 5.1 and 5.2 in the case of $\nusa < \nuc$. Spectra 1.3, 1.4, 2.3, and 2.4 have $\nuchat^\ast$ below $\nuc$, therefore, $\nuchat^\ast$ is not a visible break in the spectrum, and the segment above $\nuc$ is $\Hprime$ rather than $H$.

\begin{figure*}[htbp]
    \centering
    \begin{tabular}{cc}
        \centering
        \includegraphics[width=0.48\textwidth]{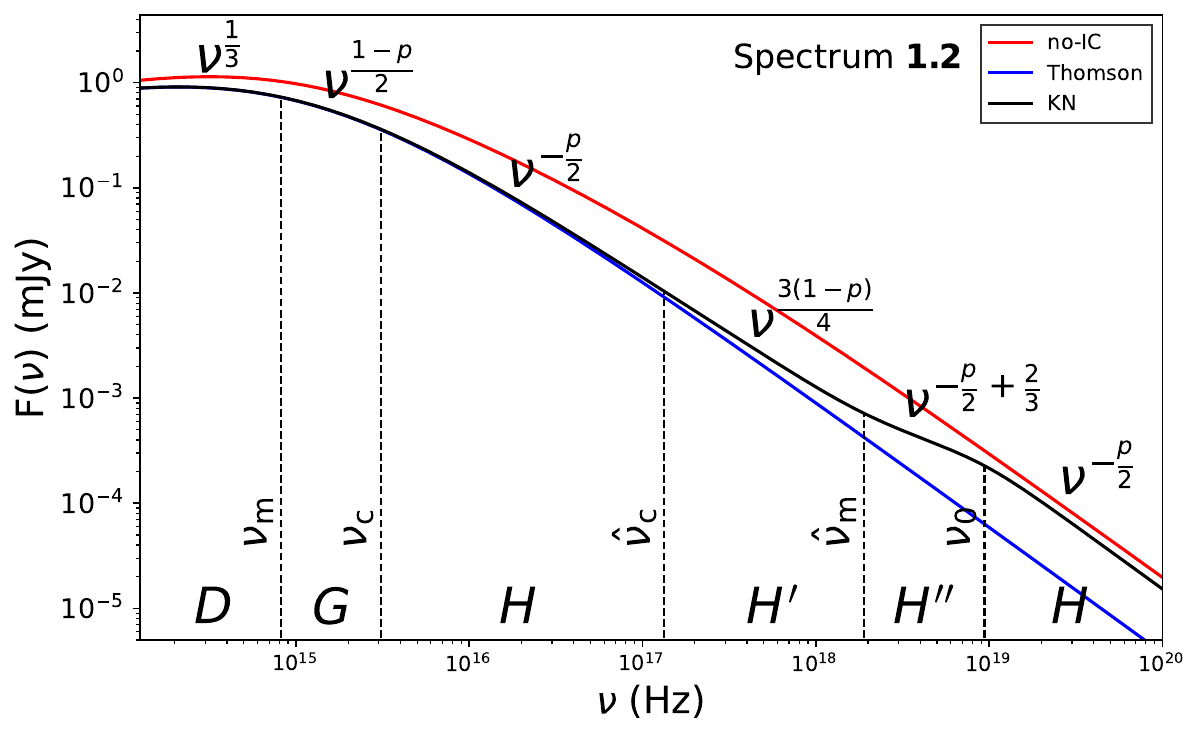} &
        \includegraphics[width=0.48\linewidth]{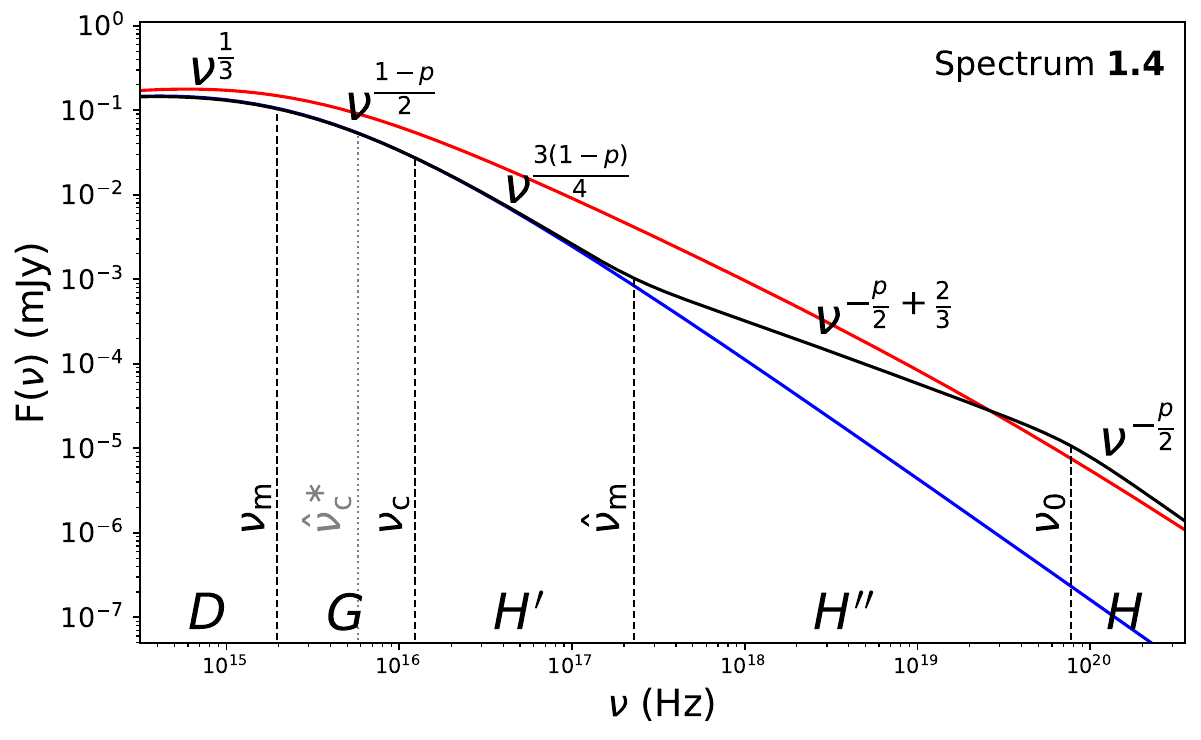}\\
        \includegraphics[width=0.48\linewidth]{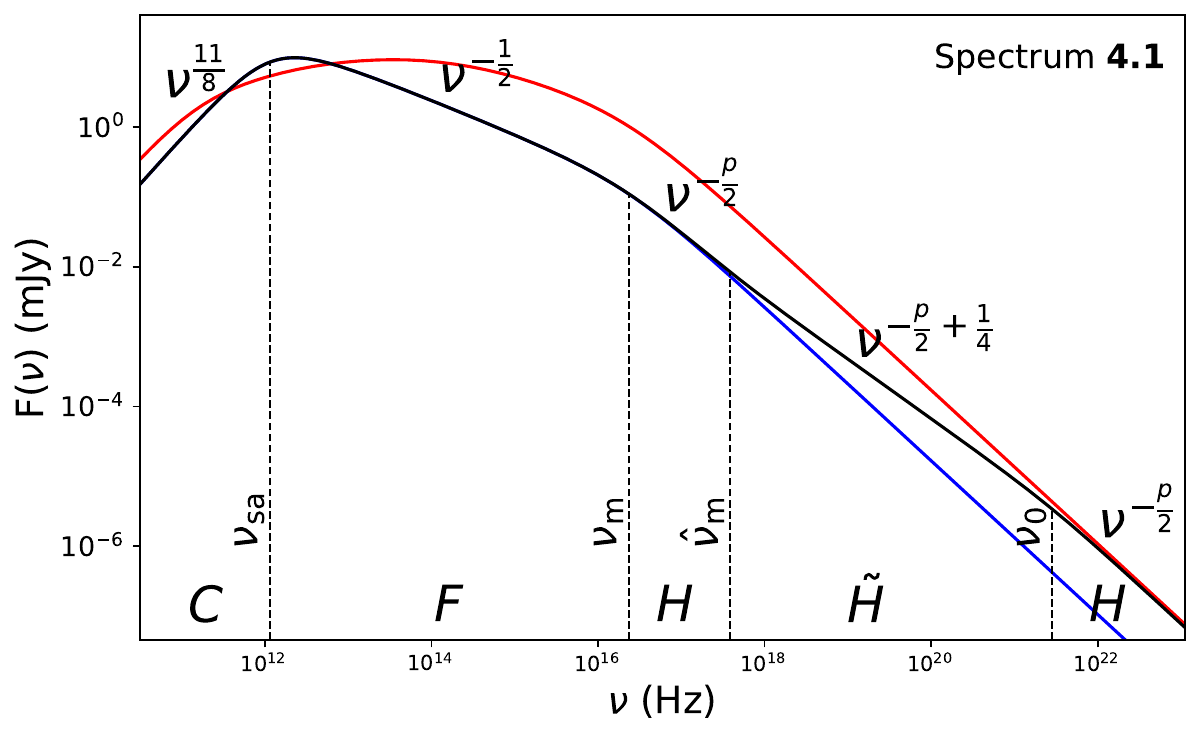}&
        \includegraphics[width=0.48\linewidth]{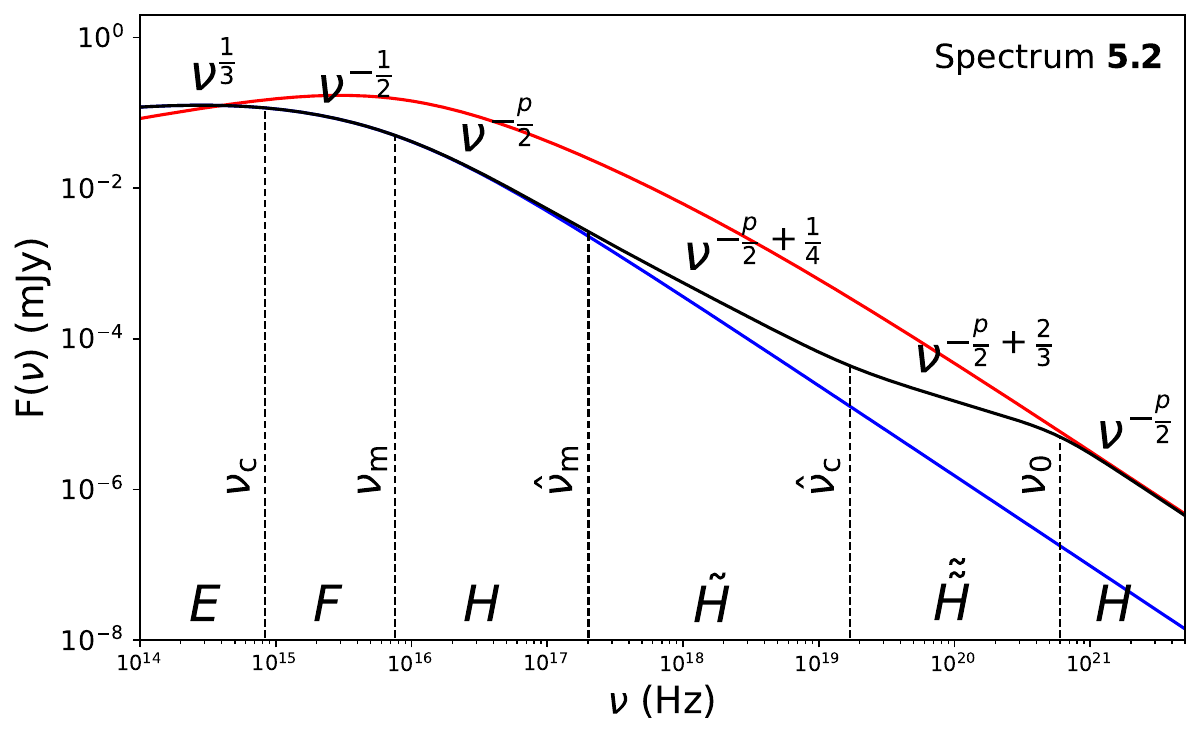}
    \end{tabular}
    \caption{Example KN spectra corresponding to the cases (and parameters) in Figure~\ref{fig:kn_diagram}. For all parameter sets, $z=1$ and $\epse=0.95$. \textbf{Case 1.2} (slow cooling).  Parameters: $k=0$, $p=2.4$, $\epsb=0.05$, $n_0=10^{-3}$, $E_{\rm 52}=30$, $t = 1$ day. \textbf{Case 1.4} (slow cooling). The $\nuchat^\ast$ break is obscured by $\nuc$ and does not result in a break on segment $G$. The KN model rises above the no-IC model at $\nunaught$ because of the smoothing used. Parameters: $k=0$, $p=2.89$, $\epsb=10^{-3}$, $n_0=10^{-3}$, $E_{\rm 52}=30, t=0.35$ days. \textbf{Case 4.1} (fast cooling). There is no change in the spectrum at $\nuchat$ which is much larger than $\nunaught$. Parameters: $k=2$, $p=2.2$, $\epsb=0.01$, $A_\ast=0.1$, $E_{\rm 52}=1$, $t=0.01$ days. \textbf{Case 5.2} (fast cooling). Parameters: $k=0$, $p=2.4$, $z=1$, $\epsb=0.03$, $n_0=1$, $E_{\rm 52}=1, t=0.02$ days. Note -- Segments $D$ and $E$ on spectra 1.2, 1.4, and 5.2 appear steeper than $\nu^{1/3}$ due to smoothing.}
    \label{fig:kn-nokn-comparison}
\end{figure*}

Spectral shape 3 applies when $\{\numax, \nuc\} < \nusa$ regardless of the relative ordering of $\numax$ and $\nuc$. Therefore, this spectrum may apply to fast and slow cooling. In all cases, KN breaks are only visible if $\nusa < \nunaught$; otherwise, the spectrum is the SOCS. Spectra 3.1-3.5 are slow-cooling; spectra 3.1 and 3.2 have $\nusa < \nuchat$, and spectrum 3.3 and 3.4 have $\nuchat < \nusa$. In each case, swapping the ordering of $\nunaught$ and $\numhat$ doubles the number of distinct spectral orderings. 
Spectrum 3.5 is the only slow-cooling ordering for shape 3 with $\numhat < \nusa$, this results in an $\Hprimeprime$ segment above $\nusa$, and $\nunaught$ is the only visible KN break. Spectra 3.6--3.10 are fast cooling spectra, Spectra 3.6 and 3.7 have $\nusa < \numhat < \nunaught$, and have alternate orderings of $\{\nuchat, \nunaught\}$. Spectra 3.8 and 3.9 have $\numhat < \nusa$ and $\nusa < \nuchat$, and in either case the ordering of $\nuchat$ and $\nunaught$ is permuted. Spectrum 3.10 has $\nuchat < \nusa$, resulting in $\Hprimeprime$ above $\nusa$ and the only visible KN break is $\nunaught$. In practice, only a subset of these spectra are expected to frequently appear in GRB afterglows, and we discuss the most prominent cases in the next section.  We demonstrate the differences between KN-corrected spectra, the SOCS, and spectra with IC cooling but without including KN effects in Figure~\ref{fig:kn-nokn-comparison}. 

\begin{deluxetable}{ccl}
\tabletypesize{\scriptsize}
\centering 
\tablecaption{KN-corrected Spectra ($\numax<\numhat$)
\label{table:kn_spectra}}
\tablehead{\colhead{No.} & \colhead{Condition} & \colhead{PLS}}
\startdata
1.1 & $\nusa<\numax<\nuc<\nuchat<\nunaught<\numhat^\ast$ & $B, D, G, H, H', H$ \\[0.2em] 
1.2 & $\nusa<\numax<\nuc<\nuchat<\numhat<\nunaught$ & $B, D, G, H, H', H'', H$ \\[0.2em] 
1.3 & $\{\nusa<\numax,\nuchat^\ast\}<\nuc<\nunaught<\numhat^\ast$ & $B, D, G, H', H$ \\[0.2em] 
1.4 & $\{\nusa<\numax,\nuchat^\ast\}<\nuc<\numhat<\nunaught$ & $B, D, G, H', H'', H$ \\[0.2em] 
\hline \\[-1.2em] 
2.1 & $\numax<\nusa<\nuc<\nuchat<\nu_0<\numhat^\ast$ & $B, A, G, H, H', H$ \\[0.2em] 
2.2 & $\numax<\nusa<\nuc<\nuchat<\numhat<\nu_0$ & $B, A, G, H, H', H'', H$ \\[0.2em] 
2.3 & $\{\numax<\nusa,\nuchat^\ast \}<\nuc<\nu_0<\numhat^\ast$ & $B, A, G, H', H$ \\[0.2em] 
2.4 & ${\{\numax<\nusa,\nuchat^\ast \}}<\nuc<\numhat<\nu_0$ & $B, A, G, H', H'', H$ \\[0.2em] 
\hline \\[-1.2em]
3.1 & $\numax<\nuc^\ast<\nusa<\nuchat<\nunaught<\numhat^\ast$ & $B, A, H, H', H$ \\[0.2em] 
3.2 & $\numax<\nuc^\ast<\nusa<\nuchat<\numhat<\nunaught$ & $B, A, H, H', H'', H$ \\[0.2em] 
3.3 & $\numax<\{\nuc^\ast,\nuchat^\ast\}<\nusa<\nunaught<\numhat^\ast$ & $B, A, H', H$ \\[0.2em]
$3.3^\ast$ & $\nuchat^\ast<\numax<\nuc^\ast<\nusa<\nunaught<\numhat^\ast$ & $B, A, H', H$ \\[0.2em]
3.4 & $\{\numax,\nuchat^\ast\}<\nuc^\ast<\nusa<\numhat<\nunaught$ & $B, A, H', H'', H$ \\[0.2em]
3.5 & $\numax<\nuchat^\ast<\nuc^\ast<\numhat^\ast<\nusa<\nunaught$ & $B, A, H'', H$ \\[0.2em]
3.6 & $\nuc^\ast<\numax<\nusa<\numhat<\nunaught<\nuchat^\ast$ & $B, A, H, \Htilde, H$ \\[0.2em]
3.7 & $\nuc^\ast<\numax<\nusa<\numhat<\nuchat<\nunaught$ & $B, A, H, \Htilde, \Htildetilde, H$ \\[0.2em]
3.8 & $\nuc^\ast<\{\numax,\numhat^\ast\}<\nusa<\nunaught<\nuchat^\ast$ & $B, A, \Htilde, H$ \\[0.2em]
3.9 & $\nuc^\ast<\{\numax,\numhat^\ast\}<\nusa<\nuchat<\nunaught$ & $B, A, \Htilde, \Htildetilde, H$ \\[0.2em]
3.10 & $\nuc^\ast<\{\numax,\numhat^\ast\}<\nuchat^\ast<\nusa<\nunaught$ & $B, A, \Htildetilde, H$ \\[0.2em]
\hline \\[-1.2em]
4.1 & $\nuac<\nusa<\numax<\numhat<\nunaught<\nuchat^\ast$ & $B, C, F, H, \Htilde, H$ \\[0.2em] 
4.2 & $\nuac<\nusa<\numax<\numhat<\nuchat<\nunaught$ & $B, C, F, H, \Htilde, \Htildetilde, H$ \\[0.2em] 
\hline \\[-1.2em]
5.1 & $\nuac<\nusa<\nuc<\numax<\numhat<\nunaught<\nuchat^\ast$ & $B, C, E, F, H, \Htilde, H$ \\[0.2em] 
5.2 & $\nuac<\nusa<\nuc<\numax<\numhat<\nuchat<\nunaught$ & $B, C, E, F, H, \Htilde, \Htildetilde, H$ \\[0.2em] 
\enddata
\tablecomments{All Klein-Nishina sub-spectra which lead to distinct spectral shapes are shown for the five spectral shapes discussed by \cite{gs02}. Break frequencies that do not introduce a new segment (such as $\hat{\nu}$ if $\nunaught < \hat{\nu}$) are labeled $\hat{\nu}^\ast$. Swapping the position of frequencies written $\{\nu_1, \nu_2\}$ does not result in a distinct spectral shape. Likewise, $\{\nu_1 < \nu_2, \nu_3\}$ implies that all orderings of $\nu_3$ relative to $\nu_1$ and $\nu_2$ have degenerate spectral shapes. 
Spectrum 3.3 and $3.3^\ast$ correspond to the same spectral shape but are written separately for clarity of notation as, in the case of $3.3^\ast$, $\nuchat < \numax$.
}
\end{deluxetable}

\section{Observational signatures of KN effects}
We now consider the observational effects of the KN corrections on GRB X-ray afterglows, for which the break frequencies regularly satisfy $\numax<\nuc<\nu_X$ (i.e., slow cooling) for a majority of the observed evolution \citep[e.g.,][]{ops+11}. All times in this section refer to time in the observer frame. In this regime, the most relevant KN spectral segments are $\Hprime$ and $\Hprimeprime$.  
The latter has a more restrictive validity condition ($\Hprimeprime$ requires $\nuchat < \numhat < \nu < \nu_0$ where $\Hprime$ only requires $\nuchat < \nu < \{\numhat, \nu_0\}$). Additionally, $\numax \propto t^{-3/2}$, which quickly pushes $\numhat$ above the X-rays ($\numhat \sim \numax^{-2} B^{-1}$). Therefore, when KN corrections are important, we expect the X-rays to be most commonly on the $\Hprime$ segment. We now derive the light curve temporal indices for both $\Hprime$ and $\Hprimeprime$ segments, considering both relative orderings of $\nuc$ and $\nuchat$ as well as both wind and ISM environments separately in each case. 

In the case that $\nuchat < \nuc$, the flux density on the $\Hprime$ segment ($\nuchat<\nuc<\nu<\nunaught$) is given by,
\begin{equation}
    F_{\Hprime}=F_m\left(\frac{\nu_c}{\nu_m}\right)^{(1-p)/2}\left(\frac{\nu}{\nu_c}\right)^{3(1-p)/4}.
\end{equation}
The cooling-break $\nuc = \nucs (1 + \Yc)^{-2}$, which for $\Yc \gg 1$ yields $\nuc \approx \nucs \Yc^{-2}$. This results in
\begin{equation}
    F_{\Hprime} \propto F_m \numax^\frac{p-1}{2} (\nucs)^\frac{p-1}{4} \Yc^\frac{1-p}{2}, 
\end{equation}
where the appropriate $\Yc$ is row 3 of Table \ref{table:Yc}.
\begin{equation}
    \Yc \propto \left[\left(\frac{\gammam}{\gammacs}\right)^{p-2} \left(\frac{\gammacs}{\gammacshat}\right)^{\frac{p-3}{2}}\right]^{\frac{2}{p-1}}
\end{equation}
The electron Lorentz factors are eliminated through $\nusyn(\gamma) \approx \Gamma \gamma^2 q_e B / (2 \pi m_e c)$, and $\nuchat$ through the relation
\begin{equation}
    \hat{\nu} \propto \Gamma^3 \nu^{-2} B
    \label{eq:nuhat_from_Gamma_and_B}
\end{equation}
resulting in
\begin{equation}
    \Yc \propto \left[\left(\frac{\numax}{\nucs}\right)^{\frac{p-2}{2}} \left(\left(\frac{\nucs}{\Gamma}\right)^3 B^{-1} \right)^{\frac{p-3}{4}}\right]^{\frac{2}{p-1}}.
\end{equation}
For slow cooling, in an ISM environment before the jet break: $\numax \propto t^{-3/2}$,  $\nucs \propto t^{-1/2}$ and  $\Gamma \propto B \propto t^{-3/8}$. Thus,
\begin{equation}
    \Yc \propto t^\frac{2-p}{p-1}
\end{equation}
and
\begin{equation}
    F_{\Hprime} \propto t^{-(3p+1)/8}.
\end{equation}
In a wind environment before the jet break $F_m \propto t^{-1/2}$, $\numax \propto t^{-3/2}$,  $\nucs \propto t^{1/2}$,  $\Gamma \propto t^{-1/4}$ and $B \propto t^{-3/4}$. Therefore,
\begin{equation}
    \Yc \propto t^\frac{p + 1}{2(1 - p)}
\end{equation}
and
\begin{equation}
    F_{\Hprime} \propto t^{3(1-p)/8}.
\end{equation}

\begin{figure*}
    \centering
    \begin{tabular}{cc}
    \includegraphics[width=\columnwidth]{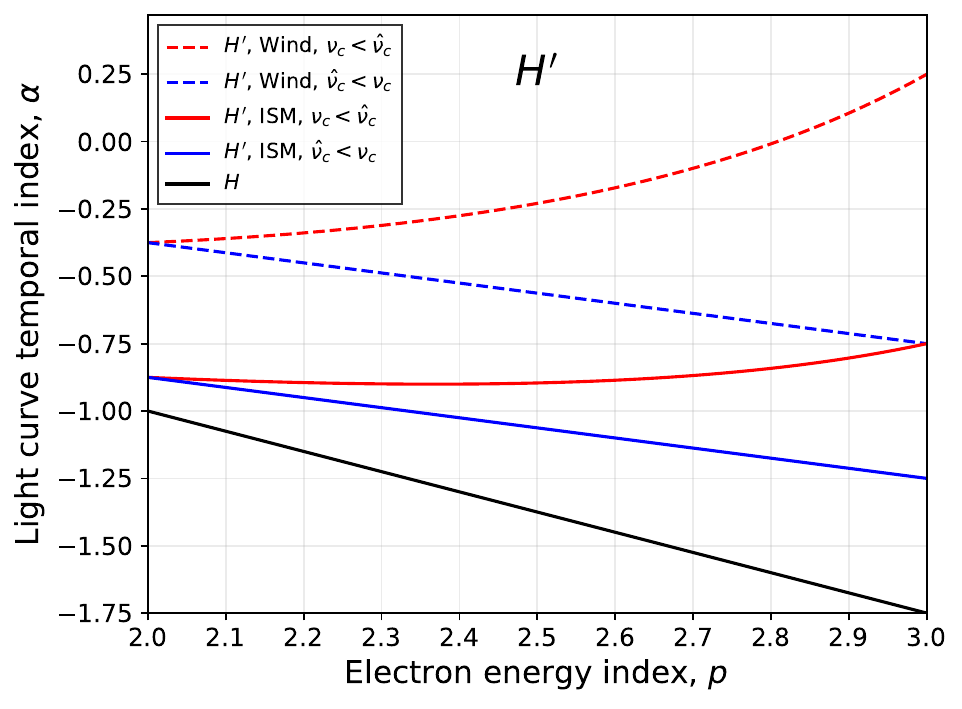} 
    \includegraphics[width=\columnwidth]{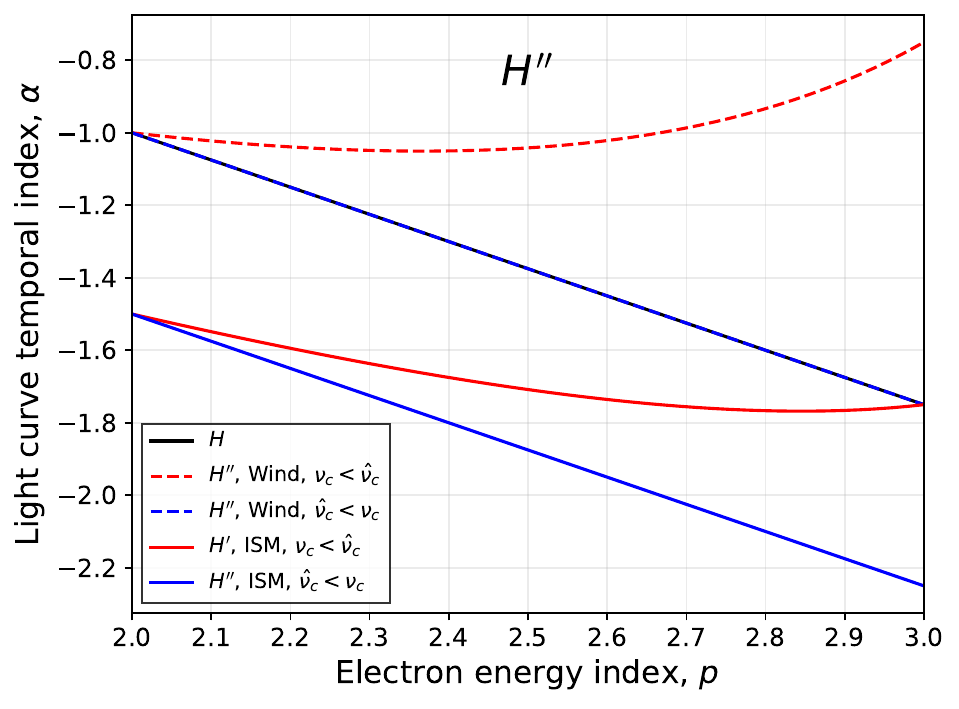}
    \end{tabular}
    \caption{Dependence of $\alpha$ ($F_\nu \propto t^\alpha$) on $p$ in wind (red) and ISM (blue) environments with $\nuc < \nuchat$ (solid) and $\nuchat < \nuc$ (dashed), on the $\Hprime$ ($\beta=3(1-p)/4$; left) and $\Hprimeprime$ ($\beta=-p/2+2/3$; right) segments, compared with segment $H$ (black). $\Yc$ decreases rapidly with time for large $p$, which can cause $\nuc$ to rise, resulting in a brief brightening if $\Yc \gg 1$. On the $\Hprimeprime$ segment, the light curve in the $\nuchat<\nuc$ regime in the wind environment (blue, dashed) is identical to that on the $H$ segment.}
    \label{fig:h_prime_time_deps}
\end{figure*}

In the regime, $\numax<\nuc<\nuchat<\nu$, the flux density is given by,
\begin{equation}
    F_{\Hprime}=F_m\left(\frac{\nuc}{\numax}\right)^{(1-p)/2}\left(\frac{\nuchat}{\nuc}\right)^{-p/2}\left(\frac{\nu}{\nuchat}\right)^{3(1-p)/4}, 
\end{equation}
For $\nuc < \nuchat$, $\Yc = \YT$, which implies
\begin{equation}
    F_{\Hprime} \propto F_m \numax^{\frac{p-1}{2}} (\nucs)^{\frac{4-p}{2}} B^{\frac{p-3}{4}} \Gamma^{\frac{3(p-3)}{4}} \YT^{p-5}.
\end{equation}
In slow cooling, for $\gammam\ll\gammac$ and $\YT \gg 1$, we have (from row 4 of Table~\ref{table:Yt}),
\begin{equation}
    \YT \approx \left(\frac{\gammam}{\gammacs}\right)^{(p-2)/(4-p)}.
\end{equation}
Therefore, in an ISM environment,
\begin{equation}
    \YT \propto t^\frac{2-p}{2(4-p)},
\end{equation}
and
\begin{equation}
    F_{\Hprime} \propto t^{\frac{7(1-p)}{8} + \frac{(2-p)(p-5)}{2(4-p)}}.
\end{equation}
Whereas, in a wind environment
\begin{equation}
    \YT \propto t^\frac{2-p}{4-p},
\end{equation}
and
\begin{equation}
    F_{\Hprime} \propto t^{\frac{(19 - 11p)}{8} + \frac{(2-p)(p-5)}{(4-p)}}.
\end{equation}

We next consider the \Hprimeprime\ segment. In the case of $\nuchat < \nuc$ the flux density in the regime $\nuchat<\nuc<\numhat<\nu$ is given by,
\begin{equation}
    F_{\Hprimeprime}=F_m\left(\frac{\nu_c}{\nu_m}\right)^{\frac{1-p}{2}}\left(\frac{\hat{\nu}_m}{\nu_c}\right)^{\frac{3(1-p)}{4}}\left(\frac{\nu}{\hat{\nu}_m}\right)^{-\frac{p}{2} + \frac23},
\end{equation}
and when $\nuc<\nuchat<\numhat<\nu$,
\begin{align}
    F_{H''}=F_m\left(\frac{\nu_c}{\nu_m}\right)^{\frac{1-p}{2}}\left(\frac{\hat{\nu}_c}{\nu_c}\right)^{-\frac{p}{2}}\left(\frac{\hat{\nu}_m}{\hat{\nu}_c}\right)^{\frac{3(1-p)}{4}} 
    \left(\frac{\nu}{\hat{\nu}_m}\right)^{-\frac{p}{2} + \frac23}.
\end{align}
Therefore, $F_{\Hprimeprime} \propto F_{\Hprime} \hat{\nu}_m^\frac{1 - 3p}{12}$ in both configurations. From equation \ref{eq:nuhat_from_Gamma_and_B}, we find $\numhat \propto t^\frac32$ before the jet break in both ISM and Wind environments. This yields the following time dependence for $F_{\Hprimeprime}$,
\begin{equation}
F_{\Hprimeprime} \propto
\begin{cases}
t^{-\frac{3p}{4}}, & \nuchat<\nuc<\numhat<\nu \\
t^{\frac{4 - 5p}{4} + \frac{(2-p)(p-5)}{2(4-p)}}, & \nuc<\nuchat<\numhat<\nu
\end{cases}
\end{equation}
in the ISM environment, and 
\begin{equation}
F_{\Hprimeprime} \propto
\begin{cases}
t^\frac{2 - 3p}{4}, & \nuchat<\nuc<\numhat<\nu \\
t^{\frac{10 - 7p}{4} + \frac{(2-p)(p-5)}{4-p}}, & \nuc<\nuchat<\numhat<\nu
\end{cases}
\end{equation}
in the wind environment. 
We summarize these results in Figure~\ref{fig:h_prime_time_deps}. 

\begin{figure}
    \centering
    \vspace{1em}
    \includegraphics[width=\columnwidth]{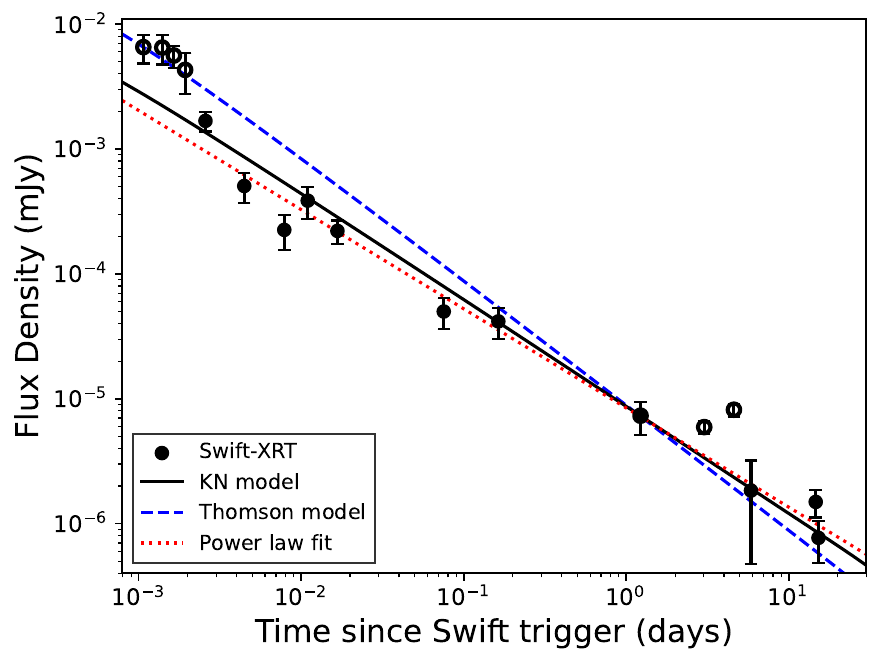}
    \caption{Swift/XRT light curve (black, markers) for GRB~210726A, together with a KN model (black, solid), an IC-cooling model without KN corrections (a ``Thomson model''; blue, dashed) and a power law fit ($t^{-0.79}$). The KN model is a better fit to the  X-ray observations than the Thomson model.}
    \label{fig:210726AXray}
\end{figure}

An example application of these results is that of SGRB~210726A, for which \cite{srl+23} found evidence of spectral and temporal signatures consistent with KN corrections associated with the $\Hprime$ segment. They present an analysis of the afterglow including radio, optical and X-ray observations and implement the exact KN model described in this paper. For their best-fit parameters of $k = 0$ (ISM), $p = 2.04$, $\epse = 0.9$, $\epsb = 1.1 \times 10^{-4}$, $E_{\rm K, iso}=8.1 \times 10^{52}$ erg, $n_0 = 7.4 \times 10^{-2}$ $\rm cm^{-3}$, the afterglow SED is in slow cooling and spectrum $1.3$ ($\numax < \nuchat < \nuc < \nu_X < \numhat < \nu_0$) spans from $4 \times 10^{-4}$ days to $400$ days. Consequently, the X-ray band is on the $\Hprime$ segment during the entire evolution (Figure~\ref{fig:210726AXray}). The spectral index of $\beta_{\rm X}=3(1-p)/4 = -0.78$ predicted in this regime is consistent with the observed value of $\beta_{\rm X}=-0.8\pm0.1$. Furthermore, the predicted light curve decay rate of $\alpha_{\rm X}=-(3p+1)/8\approx-0.89$ is much closer to the observed value of $\alpha=-0.79\pm0.04$ (from fitting a power law to the data after $2\times10^{-3}$ and excluding the late-time X-ray flare between $\approx3$--4.6~days) than the value of $\alpha_{\rm X}=(2-3p)/4\approx-1.03$ expected on segment $H$ for this value of $p$. Thus, KN corrections provide one possible solution to the observed hard X-ray spectral index and shallow X-ray light curve in this case.

\section{Summary and Conclusions}
We have provided an end-to-end description of the steps needed to calculate KN corrections to the synchrotron spectrum in the weak KN regime. Our work combines the comprehensive computation of the $Y$-parameter in the Thomson regime from \JBH\ with the detailed discussion of KN corrections from \NAS\ into a single reference that can be used to analytically compute KN-corrected synchrotron spectra of GRB afterglows. In addition to the relevant expressions for the $Y$-parameter in the Thomson regime (\YT) and the KN-corrected $Y$-parameter ($Y(\gammae)$), we also calculate (i) the value of the $Y$-parameter at $\nuc$ ($\Yc$); (ii) the expected location of the cooling break; and (iii) the location of $\nunaught$, where the effects of SSC cooling become negligible and the spectrum returns to synchrotron-dominated cooling. Our expressions for $\nuc$ and $\nunaught$ in this regime are somewhat different from those derived by previous authors. For ease of implementation and adoption, we release a Python codebase for computing each of the above. 

We have applied these results to investigate the spectral shapes expected under all possible orderings of the synchrotron and KN break frequencies, and have summarized all distinct spectral shapes expected in the presence of KN corrections.
We have additionally derived the temporal signatures of the weak KN regime in slow cooling (specifically, the variation of the light curve decay index, $\alpha$, with $p$). The corresponding expressions for $\alpha$ and $\beta$ imply new closure relations that may be useful for modeling the high-energy (in particular, X-ray) light curves of GRB afterglows. Finally, we have demonstrated the importance of KN corrections in the $p>2$ regime using the example of the short GRB 210726A. 

Our work has two immediate caveats. Like \JBH, we focus on the weak KN regime. In the strong KN regime ($\gammamhat<\gammam$), the spectrum returns to synchrotron-dominated cooling in the slow cooling case (\NAS); however, the fast strong case is more complex as the normalization of $Y(\gammae)<\min{(\gammamhat,\gammachat)}$ can no longer be straightforwardly related to $\YT$. We defer an in-depth discussion of the strong KN regime to future work. 
Second, our work explicitly focuses on the impact of KN corrections to the shape of the synchrotron spectrum. Given the knowledge of $Y(\gammae)=P_{\rm SSC}/P_{\rm syn}$, it is possible to follow previous analyses (e.g., \NAS\ and \citealt{se01}) to incorporate a discussion of the shape and normalization of the SSC spectrum itself. Such an extension would allow this prescription to be applied to afterglow observations at even higher (e.g., GeV to TeV) energies, expected to be increasingly relevant in the upcoming era of sensitive $\gamma$-ray facilities like the Cerenkov Telescope Array.

\section*{Acknowledgments}
The authors would like to thank Hendrik van Eerten, Patricia Schady, Paz Beniamini, Taylor Jacovich, and Coleman Rohde for helpful discussions. GAM acknowledges support from the UK Royal Astronomical Society. The authors thank the referee for their careful read of our manuscript and for providing useful suggestions.

\appendix

\section{The Compton Y-parameter in the Thomson regime}
\label{appendix:YT}
In the Thomson regime, KN effects are unimportant and the entire electron distribution can upscatter synchrotron photons. In this limit, we follow SE01 to obtain an approximate value for \YT\ using, 
\begin{equation}
    \frac{P_{\rm SSC}}{P_{\rm syn}} \approx \frac{U_{\rm syn}}{U_{\rm B}},
\label{eq:ic_power_sync_energy_approximation}
\end{equation}
which yields, 
\begin{equation}
    \YT(1 + \YT) = \frac{\eta \epsilon_e}{\epsilon_B}, 
\label{eq:SE_yt}
\end{equation}
where $\eta = 1$ for $\gammac<\gammam$ and $\eta=(\gammac/\gammam)^{2-p}$ for $\gammam<\gammac$.
Equation \ref{eq:SE_yt} can be solved for \YT\ using the quadratic formula or by taking the ultra-fast-cooling limit $(\gammac \ll \gammam)$, leading to
\begin{equation}
    \YT = 
    \begin{cases}
        \sqrt{\frac{\epse}{\epsb}}\eta & (\YT \gg 1) \\
        \frac{\epse}{\epsb}\eta & (\YT \ll 1), \\
    \end{cases}
\end{equation}

Incorporating the full electron distribution results in some modifications to the above estimate of \YT, which \JBH\ demonstrate by beginning with the following equivalent definition of $Y$,
\begin{equation}
    Y_T=\frac{4}{3}\sigma_T n_0 \Delta R \gammaerms
    \label{eq:YT_JBH}
\end{equation}
where $\sigma_T$ is the Thomson scattering cross-section, $n_0$ is the electron number density in the fluid rest frame, $\Delta R$ is the width of a thin shell of emitting material at the shock-front, and $\gammaerms$ is the second moment of the electron Lorentz factor distribution,
\begin{equation}
\label{eq:rms_gammae}
    \gammaerms = \frac{1}{n_0}\int_1^\infty \frac{dn_0}{d\gammae} \gammae^2 d\gammae.
\end{equation}
Evaluating equations \ref{eq:YT_JBH} and \ref{eq:rms_gammae} in each cooling regime\footnote{Using equation~\ref{eq:YT_JBH} to derive expressions for \YT\ requires the relation, $\frac{4}{3}\sigma_T n_0 \Delta R = \frac{p-2}{p-1}\frac{\epse}{\epse}\frac{1}{\gammam\gammacs}$ \citep{vaneerten2015}.}
 then yields expressions for \YT. 
For completeness, we provide the resulting expressions for \YT\ derived by \JBH\ in Table \ref{table:Yt} and summarize the key details of our implementation here. 

The fast cooling solution is a cubic polynomial which we solve using Cardano's formula, allowing for efficient computation of $Y_{\rm T, fast}$. However, $Y_{\rm T, slow}$ cannot be solved closed-form. Therefore, we use the approximation $Y_{\rm T,slow}$ as derived by \cite{bnd15} in the ultra-slow cooling limit ($\gammam \ll \gammac)$. We smooth the $\YT \gg 1$ and $\YT \ll 1$ asymptotic forms of this solution together $Y_{\rm T,slow,approx} = (Y_{\rm T,slow,approx,5}^{\mu_1} + Y_{\rm T,slow,approx,6}^{\mu_1})^{1/{\mu_1}}$ (the numerical subscripts correspond to row numbers in Table \ref{table:Yt}). Here, $\mu_1 = -1.7$ is a smoothing constant\footnote{\JBH\ describe this smoothing process but do not report the smoothing constant they employ.}.

\begin{table*}
\caption{Equations for $Y_T$}
\centering 
\begin{tabular}{c c l} 
\hline\hline 
Condition 1 & Condition 2 & \YT\ \\ [2ex] 
\hline 
$\gammac < \gammam$ & & $Y_T(1 + Y_T) = \frac{(p-2)}{(p-1)} \frac{\epse}{\epsb}\left[\frac{(p-1)}{(p-2)}(1 + Y_T) - \frac{\gammacs}{\gammam}\right]\left[(1 + Y_T)-\frac{p-1}{p}\frac{\gammacs}{\gammam}\right]^{-1}$ \\ [1.5em]
$\gammam = \gammac$ & & $Y_T = \frac{1}{2}\left(\sqrt{1 + \frac{4p}{p-1}\frac{\epse}{\epsb}} - 1\right)$ \\ [1.5em] 
$\gammam < \gammac$ & & $Y_T(1 + Y_T)^2 = {p\left[\frac{\epse}{\epsb}\frac{\gammam}{\gammacs}(1 + Y_T)^{3-p}\frac{p-2}{p-3} + \frac{\epse}{\epsb}\frac{1}{3-p}\left(\frac{\gammam}{\gammacs}\right)^{p-2}\right]}{\left[p(1+Y_T)^{1-p} - \left(\frac{\gammam}{\gammacs}\right)^{p-1}\right]}^{-1}$ \\ [1.5em]
 $\gammam \ll \gammac$ & & $\YT(1 + \YT)^{3-p} = \frac{\epse}{\epsb}\frac{1}{3-p}\left(\frac{\gammam}{\gammacs}\right)^{p-2}$ \\ [1.5em]
  $\gammam \ll \gammac$ & $\YT \gg 1$ & $\YT \approx \left[\frac{\epse}{\epsb}\frac{1}{3-p}\left(\frac{\gammam}{\gammacs}\right)^{p-2}\right]^{\frac{1}{4-p}}$ \\ [1.5em] 
  $\gammam \ll \gammac$ & $\YT \ll 1$ & $\YT \approx \frac{\epse}{\epsb}\frac{1}{3-p}\left(\frac{\gammam}{\gammacs}\right)^{p-2}$ \\ [1.5em]
\hline
\end{tabular}
\label{table:Yt}
\end{table*}

\begin{table*}
\caption{Glossary of Physical Quantities and their Definitions}
\centering 
\begin{tabular}{c l} 
\hline\hline 
Symbol & Definition \\ [2ex]
\hline
\EKiso & Isotropic equivalent kinetic energy \\
k & Power-law index of radial density profile, $\rho\propto r^{-k}$\\
$n_0$ & Density in particles/cm$^3$ for a uniform-density environment\\
$\Astar$ & Wind mass-loss parameter ($\rho=Ar^{-2}$ with $A=5\times10^{11}\Astar\,{\rm g\,cm}^{-1}$; see \citealt{cl99})\\
$\thetajet$ & Jet opening half-angle\\
\epse & Fraction of shock energy given to relativistic electrons\\
\epsb & Fraction of shock energy given to magnetic fields\\
$z$ & Cosmological redshift\\
$\gammae$ & Electron Lorentz factor\\
$\frac{dn_0}{d\gammae}$ & Number density of electrons with Lorentz factor, $\gammae$\\
$p$ & Power-law index of electron energy distribution, $\frac{dn_0}{d\gammae} \propto\gamma_e^{-p}$\\
$\gammam$ & Minimum energy of electron injection Lorentz factor distribution\\
$\gammacs$ & Electrons with $\gammae>\gammacs$ are cooling over the age of the system due to synchrotron radiation\\
$\gammac$ & Electrons with $\gammae>\gammac$ are cooling over the age of system due to synchrotron+SSC losses\\
$Y$ & Compton $Y$-parameter, most generally referring to $Y(\gammae)$\\
$\YT$ & Compton $Y$ in the Thomson regime, independent of $\gammae$\\
$\Yc$ & Compton $Y(\gammae)$ at $\gammae=\gammac$\\
$\gammanaught$ & Electron Lorentz factor where  $Y(\gammae=\gammanaught)=1$\\
$\Gamma$ & Bulk Lorentz factor of the post-shock fluid\\
$B$ & Magnetic field strength in the post-shock fluid rest frame\\
%$m_e$ & Electron mass\\
%$c$ & Speed of light\\
$\sigma_T$ & Electron Thomson cross section\\
$\nusyn(\gammae)$ & Characteristic synchrotron frequency of $\gammae$ electrons\\
$\numax$ & Characteristic synchrotron frequency of $\gammam$ electrons\\
$\nucs$ & Characteristic synchrotron frequency of $\gammacs$ electrons\\
$\nuc$ & Characteristic synchrotron frequency of $\gammac$ electrons\\
$\nunaught$ & Frequency above which SSC cooling is unimportant\\
$\nusa$ & Frequency below which radiation is synchrotron self-absorbed\\
$\gammaehat$ & The maximum-energy electrons capable of scattering $\gammae$ electrons due to KN effects\\
$\gammaetilde$ & Electrons responsible for maximum-energy radiation that can be IC scattered by $\gammae$ electrons\\
$\nusyn^{\rm peak}$ & Frequency where $\nusyn F_{\nu,\rm syn}$ peaks\\
SOCS & Synchrotron spectrum in the absence of IC \& KN effects\\
$\gammamhat$ & $\gammaehat$ for $\gammae = \gammam$\\
$\gammachat$ & $\gammaehat$ for $\gammae = \gammac$\\
$\gammacshat$ & $\gammaehat$ for $\gammae = \gammacs$\\
$\numhat$ & Characteristic synchrotron frequency of $\gammamhat$ electrons\\
$\nuchat$ & Characteristic synchrotron frequency of $\gammachat$ electrons\\
$\beta$ & Power-law index of observed radiation spectrum, $F_\nu\propto\nu^\beta$\\
$\alpha$ & Power-law index of observed light curves, $F_\nu\propto t^\alpha$\\
\hline
\end{tabular}
\label{tab:glossary}
\end{table*}

\bibliographystyle{aasjournal}

\bibliography{knrefs,grb_alpha}
\end{document}

%% file: defs.tex
\hypersetup{colorlinks=true, urlcolor=blue, citecolor=blue}

\newcommand{\EKiso}{\ensuremath{E_{\rm K,iso}}}

\newcommand{\epse}{\ensuremath{\epsilon_{\rm e}}}
\newcommand{\epsilone}{\ensuremath{\epsilon_{\rm e}}}
\newcommand{\epsb}{\ensuremath{\epsilon_{\rm B}}}
\newcommand{\epsilonb}{\ensuremath{\epsilon_{\rm B}}}
\newcommand{\dens}{\ensuremath{n_{0}}}
\newcommand{\Astar}{\ensuremath{A_{*}}}

\newcommand{\thetajet}{\ensuremath{\theta_{\rm jet}}}

\newcommand{\p}{\ensuremath{^{\prime}}}

\newcommand{\JBH}{JBH21}
\newcommand{\NAS}{NAS09}

\newcommand{\nusyn}{\nu_{\rm syn}}

\newcommand{\nusa}{\ensuremath{\nu_{\rm sa}}}
\newcommand{\nuac}{\ensuremath{\nu_{\rm ac}}}
\newcommand{\numax}{\ensuremath{\nu_{\rm m}}}
\newcommand{\nuc}{\ensuremath{\nu_{\rm c}}}

\newcommand{\gammae}{\ensuremath{\gamma_{\rm e}}}
\newcommand{\gammastar}{\ensuremath{\gamma^*}}
\newcommand{\nucs}{\ensuremath{\nu_{\rm c}^{\rm s}}}
\newcommand{\gammanaught}{\ensuremath{\gamma_{\rm 0}}}
\newcommand{\nunaught}{\ensuremath{\nu_{\rm 0}}}

\newcommand{\Htilde}{\ensuremath{\tilde{H}}}
\newcommand{\Htildetilde}{\ensuremath{\tilde{\tilde{H}}}}

\newcommand{\Hprime}{\ensuremath{H^\prime}}
\newcommand{\Hprimeprime}{\ensuremath{H^{\prime\prime}}}

\newcommand{\gammaerms}{\ensuremath{\langle \gamma_{\rm e}^2 \rangle}}

\newcommand{\gammam}{\ensuremath{\gamma_{\rm m}}}
\newcommand{\gammac}{\ensuremath{\gamma_{\rm c}}}
\newcommand{\gammacs}{\ensuremath{\gamma_{\rm c}^{\rm s}}}
\newcommand{\gammacshat}{\ensuremath{\hat{\gamma}_{\rm c}^{\rm s}}}

\newcommand{\gammamhat}{\ensuremath{\hat{\gamma}_{\rm m}}}
\newcommand{\gammaetilde}{\ensuremath{\tilde{\gamma}_{\rm e}}}
\newcommand{\gammaehat}{\ensuremath{\hat{\gamma}_{\rm e}}}
\newcommand{\Yc}{\ensuremath{Y_{\rm c}}}
\newcommand{\YT}{\ensuremath{Y_{\rm T}}}
\newcommand{\gammachat}{\ensuremath{\hat{\gamma}_{\rm c}}}
\newcommand{\gammaself}{\ensuremath{\gamma_{\rm self}}}

\newcommand{\nuchat}{\ensuremath{\hat{\nu}_{\rm c}}}
\newcommand{\numhat}{\ensuremath{\hat{\nu}_{\rm m}}}